\documentclass{aa}
\pdfoutput=1
\usepackage[utf8]{inputenc}
\usepackage{bm}
\usepackage{float}
\usepackage{epsf,graphics,graphicx}
\usepackage[varg]{txfonts}
\def\lsim{\!\!\!\phantom{\le}\smash{\buildrel{}\over
 {\lower2.5dd\hbox{$\buildrel{\lower2dd\hbox{$\displaystyle<$}}\over
                                 \sim$}}}\,\,}
\def\gsim{\!\!\!\phantom{\ge}\smash{\buildrel{}\over
{\lower2.5dd\hbox{$\buildrel{\lower2dd\hbox{$\displaystyle>$}}\over
                               \sim$}}}\,\,} 
\def\asec{\ifmmode ^{\prime\prime}\else$^{\prime\prime}$\fi}

\begin{document}

\title{LOFAR HBA Observations of the Euclid Deep Field North (EDFN)}
\titlerunning{LOFAR observations of the EDFN}
\author{M.~Bondi\inst{1}\thanks{\email{marco.bondi@inaf.it}}
        \and R.~Scaramella\inst{2,3}
        \and G.~Zamorani\inst{4}
        \and P.~Ciliegi\inst{4}
        \and F.~Vitello\inst{1,5}
        \and M.~Arias\inst{6}
        \and P.N.~Best\inst{7}
        \and M.~Bonato\inst{1}
        \and A.~Botteon\inst{1}
        \and M.~Brienza\inst{4,8}
        \and G.~Brunetti\inst{1}
        \and M.J.~Hardcastle\inst{9}
        \and M.~Magliocchetti\inst{10}
        \and F.~Massaro\inst{11}
        \and L.K~Morabito\inst{12,13}
        \and L.~Pentericci\inst{3}
        \and I.~Prandoni\inst{1}
        \and H.J.A.~R\"ottgering\inst{6}
        \and T.W.~Shimwell\inst{14,6}
        \and C.~Tasse\inst{15,16}
        \and R.J.~van Weeren\inst{6}
        \and G.J.~White\inst{17,18}
        }
        \authorrunning{M. Bondi et al.}
        
\institute{INAF - Istituto di Radioastronomia, Via Gobetti 101, 40129, Bologna, Italy
\and INFN-Sezione di Roma, Piazzale Aldo Moro, 2 - c/o Dipartimento di
Fisica, Edificio G. Marconi, I-00185 Roma, Italy
\and INAF - Osservatorio Astronomico di Roma, Via Frascati 33, I-00078
Monteporzio Catone, Italy
  \and INAF - Osservatorio di Astrofisica e Scienza dello Spazio di Bologna, Via
 Gobetti 93/3, 40129, Bologna, Italy
 \and  INAF - Osservatorio Astrofisico di Catania, Via Santa Sofia 78, 95123, Catania, Italy)
 \and Leiden Observatory,Leiden University, POBox 9513, 2300RA Leiden, The Netherlands
 \and Institute for Astronomy, University of Edinburgh, Royal Observatory, Blackford Hill, Edinburgh, EH9 3HJ, UK 
 \and Dipartimento di Fisica e Astronomia, Università di Bologna, via P. Gobetti 93/2, I-40129, Bologna, Italy
  \and Centre for Astrophysics Research, University of Hertfordshire, College Lane, Hatfield AL10 9AB, UK 
\and INAF - IAPS, Via Fosso del Cavaliere 100, 00133, Rome, Italy 
\and  Dipartimento di Fisica, Universit\'a degli Studi di Torino, via Pietro Giuria 1, I-10125 Torino, Italy
 \and 
 Centre for Extragalactic Astronomy, Department of Physics, Durham University, Durham DH1 3LE, UK 
\and Institute for Computational Cosmology, Department of Physics, University of Durham, South Road, Durham DH1 3LE, UK 
\and 
ASTRON, Netherlands Institute for Radio Astronomy, Oude Hoogeveensedijk 4, 7991 PD, Dwingeloo, The Netherlands 
\and GEPI \& ORN, Observatoire de Paris, Université PSL, CNRS, 5 Place Jules Janssen, 92190 Meudon, France 
\and Department of Physics \& Electronics, Rhodes University, PO Box 94, Grahamstown, 6140, South Africa 	
\and School of Physical Sciences, The Open University, Walton Hall, Milton Keynes, MK7 6AA, UK
\and RAL Space, STFC Rutherford Appleton Laboratory, Chilton, Didcot, Oxfordshire, OX11 0QX, UK 
}

\abstract{We present the first deep (72 hours of observations) radio image of the Euclid Deep Field North (EDFN) obtained with the LOw-Frequency ARray (LOFAR) High Band Antenna (HBA) at 144 MHz. The EDFN is the latest addition to the
LOFAR Two-Metre Sky Survey (LoTSS) Deep Fields and these observations represent the first data release for this field. 
The observations produced a $6\asec$ resolution image with a central r.m.s. noise of $32\,\mu$Jy\,beam$^{-1}$.  A catalogue of $\sim 23,000$ radio sources above a signal-to-noise ratio (SNR) threshold of 5 is extracted from the inner circular 10 deg$^2$ region.
We discuss the data analysis and we provide a detailed description of how we derived the catalogue of radio sources and  on the issues related to direction-dependent calibration and their effects on the final products. Finally, we derive the radio source counts at 144 MHz in the EDFN using catalogues of mock radio sources to derive the completeness correction factors. The source counts in the EDFN are consistent with those obtained from the first data release of the other LoTSS Deep Fields (ELAIS-N1, Lockman Hole and Bootes),
despite the different method adopted to construct the final catalogue and to assess its completeness.
}

\maketitle

\section{Introduction}
In the last twenty years it has been demonstrated that panchromatic surveys (from X-rays to radio wavelengths) of selected regions of the sky
\citep[e.g.][]{2005Msngr.119...30L, 2007ApJS..172....1S, 2011MNRAS.413..971D, 2011ApJS..197...35G, 2016ApJ...820...82C,
2018A&A...620A.152F, 2018A&A...616A.174P, 2022ApJS..258...11W, 
2022ApJ...935..110T, 2022arXiv221105792F}
are fundamental to provide the database on which to build a consensus
on galaxy formation and evolution \citep[e.g.][]{2014ARA&A..52..291C,2015ARA&A..53...51S, 2020ARA&A..58..661F}.
Observations at different wavelengths can reveal different aspects of the structure and properties of the galaxies, such as the distribution of stars, gas, and dust, as well as the presence of active galactic nuclei (AGN). For example, optical and infrared observations can provide information on the stars within a galaxy, while radio and X-ray observations can reveal the presence of AGN and the energetic processes associated with them. By combining data at multiple wavelengths, astronomers can hence obtain a comprehensive picture of how galaxies form and evolve over time.

Observations in the radio waveband are extremely important to study the physical processes connected to star formation, the properties of supermassive black holes and the interplay between star formation and AGN across cosmic time. The 
radio continuum emission is not affected by dust extinction
and therefore provides a dust-unbiased star formation tracer \citep[e.g.][]{1992ARA&A..30..575C,
2000ApJ...544..641H, 2008MNRAS.386.1695S, 2009ApJ...690..610S,2017A&A...602A...5N, 2022ApJ...941...10V,2023MNRAS.523.6082C}. Moreover, only at radio frequencies one
can reliably identify and probe low-luminosity jet-mode AGN hosted by the most massive galaxies, and therefore investigate the effects of feedback on their growth \citep[e.g.][]{2006MNRAS.365...11C, 2006MNRAS.368L..67B,2007MNRAS.376.1849H,2009ApJ...699L..43S, 2012MNRAS.421.1569B,2014ARA&A..52..589H,2017A&A...602A...6S,2022MNRAS.513.3742K, 2022MNRAS.511.3250M}. Present and future radio surveys can reach the depth needed to detect star-forming galaxies as well as quasar-mode AGN which are typically faint in the radio band \citep{2012ApJS..203...15B,
2016A&ARv..24...13P,2023MNRAS.523.1729B}.
However, most of the existing deep radio observations required to study these source populations at high redshifts are usually limited to small regions of the sky, ranging from tens of square arcminutes to a few square degrees
\citep[e.g.][]{1998MNRAS.296..839H, 2003AJ....125..465H, 
1999MNRAS.302..222C,2000A&AS..146...41P, 2018MNRAS.481.4548P,2003A&A...403..857B,
2007A&A...463..519B, 2005AJ....130.1373H, 2007ApJS..172...46S,
2010ApJS..188..384S,2007A&A...471.1105T,2008AJ....136.1889O,
2008ApJS..179..114M,2013ApJS..205...13M,2010ApJS..188..178M,
2017A&A...602A...1S,2017ApJ...839...35M,2018ApJS..235...34O,
2020MNRAS.495.1188M,2022ApJ...941...10V,2022ApJ...924...76A, 
2022MNRAS.509.2150H, 2022A&A...668A.133D,2023MNRAS.520.2668H}.
In this context, large area  (e.g. tens of degrees) surveys down to unprecedented depths are planned with new and upgraded facilities \citep[e.g.][]
{2016mks..confE...6J}.
The LOw-Frequency ARray \citep[LOFAR,][]
{2013A&A...556A...2V}  plays a key role in this framework by combining a wide field of view with high sensitivity and angular resolution.
The recently published second data release of the LOFAR Two-Metre Sky Survey \citep[LoTSS,][]{2017A&A...598A.104S, 2019A&A...622A...1S, 2022A&A...659A...1S}  has publicly released images covering  
5,700 square degrees  in the northern sky centred at approximately 12h45m +44$^\circ$30$^\prime$ and 1h00m +28$^\circ$00$^\prime$ and spanning 4178 and 1457 square degrees, respectively. The observations are carried out at the central frequency of 144 MHz and each pointing is observed for $\sim 8$ hours. The images have a median $1\sigma$ r.m.s.  sensitivity of $83\, \mu$Jy\,beam$^{-1}$ at $6\asec$ resolution. When completed the LoTSS observations will cover the whole Northern sky. To complement the LoTSS observations, deeper pointings \citep[the LoTSS Deep Fields,][]{ 2023MNRAS.523.1729B} in regions already covered by extensive and deep multi-wavelengths ancillary
observations, are being carried out with LOFAR with the aim to reach an rms noise of $\simeq 10\,\mu$Jy\,beam$^{-1}$ over a sky area of $\simeq 50$ deg$^2$. The LoTSS Deep Fields data release 1 (DR1) accounts for about 1/3 of the integration time for 3 deep fields (Lockman Hole, Bootes and ELAIS-N1) and provides radio images and catalogues 
\citep{2021A&A...648A...1T,2021A&A...648A...2S}, near-infrared optical identifications \citep{2021A&A...648A...3K}, photometric redshifts \citep{2021A&A...648A...4D}, and host galaxies classification and properties \citep{2023MNRAS.523.1729B}.

The North Ecliptic Pole (NEP) region is the fourth field of the LoTSS Deep Fields project \citep{2023MNRAS.523.1729B}, but was not included in the first LoTSS-Deep data release as the data were obtained later. This
field was chosen because the NEP is the location of the Euclid Deep Field North (EDFN), one of the deep fields observed by the Euclid mission \citep{2022A&A...662A.112E}, and the 
only one in the Northern sky. Euclid observations will provide sub-arcsecond near-IR
imaging down to $H=26$ mag over a 20 deg$^2$ field centred at RA$=269.73$ deg and DEC$=+66.02$ deg. 
This paper complements the LoTSS Deep Fields DR1  
presenting  the LOFAR $6\asec$ resolution image at 144 MHz and the radio source catalogue obtained from the first 72 hours in the EDFN. Other publications in preparation will present the near-IR/optical identifications of the radio sources and will focus
on the results obtained with the inclusion of the LOFAR International Stations that allow improvement of the angular resolution. 
The LOFAR observations of the EDFN have been completed in summer 2023, totalling around 400 hours of observations (expected final noise $\sim 12\,\mu$Jy\,beam$^{-1}$) and
this complete dataset is now being processed and analysed.

This paper is organized as follows. In Sections 2 and 3 we describe the 72 hours duration LOFAR observations and summarise the data calibration
and imaging procedures, respectively. Section 4 contains a detailed
description of the methods used to test the reliability of the LOFAR data products, including the refinement of the amplitude calibration, the compilation of the final radio source catalogue and 
an analysis of the properties of the radio sources at different 
distances from the field centre. The generation of mock samples of realistic radio sources used to derive the completeness factors to be applied to the source counts is described in Section 5.
Finally, Section 6 presents the radio source counts obtained in the EDFN and a brief comparison with those obtained from the other LoTSS Deep Fields.

Throughout this paper we adopt the spectral index convention
$S_\nu\propto\nu^{-\alpha}$.

\section{LOFAR observations}

LOFAR observed the EDFN for 72 hours at 144 MHz during cycle 12 (proposal 
LC12\_027, P.I. van Weeren). The proposal combined EDFN observations with those of the  galaxy cluster Abell 2255 \citep{2022SciA....8.7623B}.
This was possible because the targets, EDFN and Abell 2255, are about 5 degrees apart on the sky (resulting in minimal sensitivity losses due to the LOFAR HBA tile beam), and with the adopted setup we could split the LOFAR beam to observe both targets.
The observations presented in this paper were obtained using a single pointing centre, shifted by $\simeq 30\asec$ with respect to the EDFN positions. 

The 72 hours were split over 9 nights in the period June to
November 2019 (see Table~\ref{tab:obs}). The observations were carried out with the high-frequency band
antennas in configuration HBA\_DUAL\_INNER which provides a uniform shape of the
primary beam over the whole of the LOFAR Dutch stations
(i.e only using the inner 24 tiles of the 48 tiles on the remote stations).
All data sets were recorded with an integration time of 1s,
a 48 MHz bandwidth centred at 144 MHz and a channel width
of 3.05 kHz. The data were then passed through the standard
LOFAR pre-processing pipeline \citep{2010arXiv1008.4693H} which performed the Dysco compression to reduce the data size \citep{2016A&A...595A..99O}, the RFI flagging using the AOflagger tool \citep{ 2012A&A...539A..95O} and averaged down the data to a channel width of 12.2 kHz. Time resolution remains unchanged.

All the observations were preceded and followed by a $\sim 15$ minute run on the 
calibrators 3C295 and 3C48, respectively. The latter source was selected as primary amplitude calibrator for the data sets.
During the nine epochs of observations, the number of observing stations, including the international stations, varied from 47 to 51: two nights had one international station missing and one night had 4 stations (two international and two core stations) missing. In the standard LOFAR data analysis aimed to produce a $\sim 6\asec$ angular resolution image only the data from the 38 Dutch stations (with  baseline lengths in the range 0.15--100\,km) are processed. The international stations are flagged out at an early stage of processing to reduce the sizes of intermediate data products since they are not used.

\begin{table}
\caption{EDFN Observations}
\centering
\begin{tabular}{cccc}
\hline
 SAS ID & Date & Duration & DI rms \\
        &      &   \small{(s)} & \small{(mJy\,beam$^{-1}$)} \\ 
\hline
 L720376  & 2019-06-07 20:03:59 & 29180 & 0.26 \\
 L725452  & 2019-06-22 19:00:01 & 29170 & 0.25 \\
 L726706  & 2019-06-28 18:00:01 & 30060 & 0.25 \\
 L727108  & 2019-07-03 18:00:01 & 29170 & 0.28 \\
 L728072  & 2019-07-08 17:51:01 & 29180 & 0.32 \\
 L733075  & 2019-08-09 16:30:00 & 29180 & 0.26 \\
 L746862  & 2019-09-28 12:00:01 & 29180 & 0.45 \\
 L747611  & 2019-10-04 12:41:01 & 29170 & 0.25 \\
 L751364  & 2019-11-15 09:11:00 & 29140 & 0.23 \\
\end{tabular}
\tablefoot{Col.1: unique LOFAR SAS id; Col.2: starting date and hour of the observation (format yy-mm-dd hh:mm:ss); Col.3: duration of observation in seconds; Col.4: r.m.s.
of the direction independent image obtained using a smaller bandwidth.}
\label{tab:obs}
\end{table}

\section{Data calibration and imaging}
The data calibration and imaging were performed using the OCCAM infrastructure for High
Performance Computing (HPC) run by the  Competence Centre for Scientific Computing 
(C3S), a joint interdepartmental advanced research centre of the Turin University and
the Italian National  Institute for Nuclear Physics (INFN)
\citep{17:occam:chep}.

The data were downloaded from the LOFAR Long Term Archive and copied over to the OCCAM system. The data are stored using the Dysco data compression format \citep{2016A&A...595A..99O} that allows to reduce the data volume by roughly a factor of 4. Each night of observation amounts to $\sim 4.2$ TB, including the calibrator
scans. For the data reduction we used one OCCAM Fat Node, running 48  
cores with 768 GB of RAM available. 

For the calibration and imaging we followed the same steps used for the Bootes and Lockman Hole deep 
fields and explained in details in \citet{2021A&A...648A...1T}. 
Here we briefly summarise these steps.
Direction independent calibration was performed using the
\texttt{\small PREFACTOR} version 3 pipeline \citep{2019A&A...622A...5D, 2016ApJS..223....2V,
2016MNRAS.460.2385W}\footnote{https://github.com/lofar-astron/prefactor} using 3C48 as amplitude calibrator.
The \texttt{\small PREFACTOR} calibrator pipeline was run individually for each of the nine nights and the resulting bandpasses were checked for possible antenna malfunctions. Then, the  \texttt{\small PREFACTOR} target pipeline
was run for each night producing the calibration tables to be applied in the next data reduction step.
Direction dependent calibration was performed using the
\texttt{\small DDF-pipeline} \citep{2019A&A...622A...1S, 2021A&A...648A...1T},\footnote{https://github.com/mhardcastle/ddf-pipeline}
which combines the solver \texttt{\small KillMS}\footnote{https://github.com/saopicc/killMS} \citep{2014A&A...566A.127T,2014arXiv1410.8706T,2015MNRAS.449.2668S} and the imager
\texttt{\small DDFacet}\footnote{https://github.com/saopicc/DDFacet}
\citep{2018A&A...611A..87T}.
First, each night was processed to an early stage of reduction, that stopped after producing the direction independent corrected image. These images were checked for evident issues, such as poor ionospheric conditions and processing failures. Seven nights produced images with very similar r.m.s. noise values with the remaining two
with slightly higher but still acceptable values (the noise values are reported in Tab.~\ref{tab:obs}). 
The night with the best noise (day 2019-11-15) was fully processed with \texttt{\small DDF-pipeline} to produce the sky model and,
finally, the \texttt{\small DDF-pipeline} was run combining all the nine nights together using this sky model as a starting model for the self-calibration direction-dependent stage.

\begin{figure*}[htp]
 \centering
 \includegraphics[width=18cm]{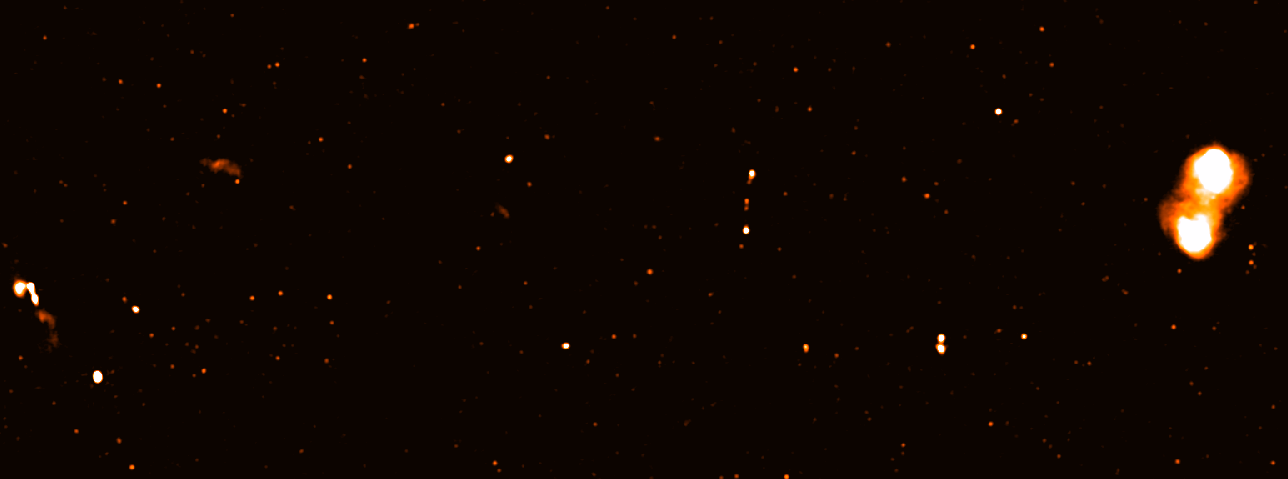}
 \caption{LOFAR image of the Euclid Deep Field North: the image shows only a sub-region $\sim 0.6\times 0.3$ deg$^2$ centred at RA=270.3086 deg and DEC=65.882 deg}
 \label{fig:lofar_sub}
\end{figure*}

\subsection{Pipeline products}

After the completion of the \texttt{\small DDF-pipeline} we obtained the deep (72 hours) full resolution (6 arcsec) Stokes I radio image centred on the EDFN. The size of the image is an input parameter for the pipeline and is usually set to $20,000\times 20,000$ pixels (1 pixel $=$ 1.5 arcsec). In processing 
the EDFN data sets we set this size to a slightly larger value ($21,500\times 21,500$ pixels, $\sim 9\times 9$ deg$^2$) to mitigate artefacts associated with a few bright sources at the edges of the field. A lower resolution Stokes I image (20 arcsec) was also produced
by the pipeline, but this image was not used as part of the analysis presented here.
The final image (before the refinement of the amplitude scale, see Sec.~\ref{sec:scale}) has a central r.m.s noise of 36 $\mu$Jy\,beam$^{-1}$.
 A $\sim 0.6\times 0.3$ deg$^2$  inset of the final image is shown in Fig.~\ref{fig:lofar_sub}.

A preliminary catalogue of the whole field was produced from the full resolution image 
using PyBDSF \citep{2015ascl.soft02007M}. PyBDSF fits individual Gaussian components in
regions selected on the basis of the local noise, extracting sources that can either be
composed of a single or multiple components. Very extended radio sources or sources whose
brightness spatial distribution can not be properly modelled by Gaussian components are effectively recovered using 
wavelets. The parameters used to run PyBDSF on the EDFN are the same used for the other Deep 
Fields and listed in Table C.1 in \citet{2021A&A...648A...2S}. PyBDSF found almost 50,000
sources above a signal-to-noise ratio (SNR) threshold of 5 over 
the whole imaged area.

\section{Data analysis}
In this section we describe the methods adopted to test the reliability of the data products, refining the amplitude calibration, producing the final catalogue of radio sources and classifying all the sources as resolved or unresolved.

\subsection{Direction-dependent effects}
\label{sec:ddeffects}
 We decided to investigate the possible effects on the catalogued source parameters introduced by the choice of  
a different field faceting geometry and of a different sky model during the direction-dependent calibration.
For this reason the EDFN datasets were independently reduced in Leiden using a different computer infrastructure. During this procedure  the image obtained from day 
2019-06-07 was adopted as sky model along with a different faceting geometry. The two final
images, the one obtained by us and the one obtained in Leiden, were then compared.
For each image we measured
the median noise in annular regions at increasing distances from the centre and 
found that the median values were consistent to within 1\% (corresponding to $\sim 0.3$
$\mu$Jy\,beam$^{-1}$) up to a radius of 2.5 deg from the field centre. Beyond this distance
the difference in the median noise increases, reaching 4\% at 4 deg from the field centre. Then, we used the same version of PyBDSF with the same set of parameters on the two images obtaining two catalogues. The two catalogues (hereafter dubbed
as Leiden catalogue and Turin catalogue) were cross-matched using a matching radius
of 3 arcsec, and only the 2918 matched sources with SNR$> 50$ in both catalogues were selected.
These sources were then split according to their distance
 $r$ from the field centre in four annular regions: $r < 1$ deg, $1 < r < 2$ deg,
 $2 < r < 3$ deg and $3 < r < 4$ deg. For each group of matched sources we calculated the ratio between  the total (peak) fluxes obtained in the Turin and Leiden catalogues, and we derived the median and the scaled median absolute deviation (MAD) in each of the four distance intervals.

\begin{figure*}[htp]
 \centering
 \includegraphics[width=6.0cm]{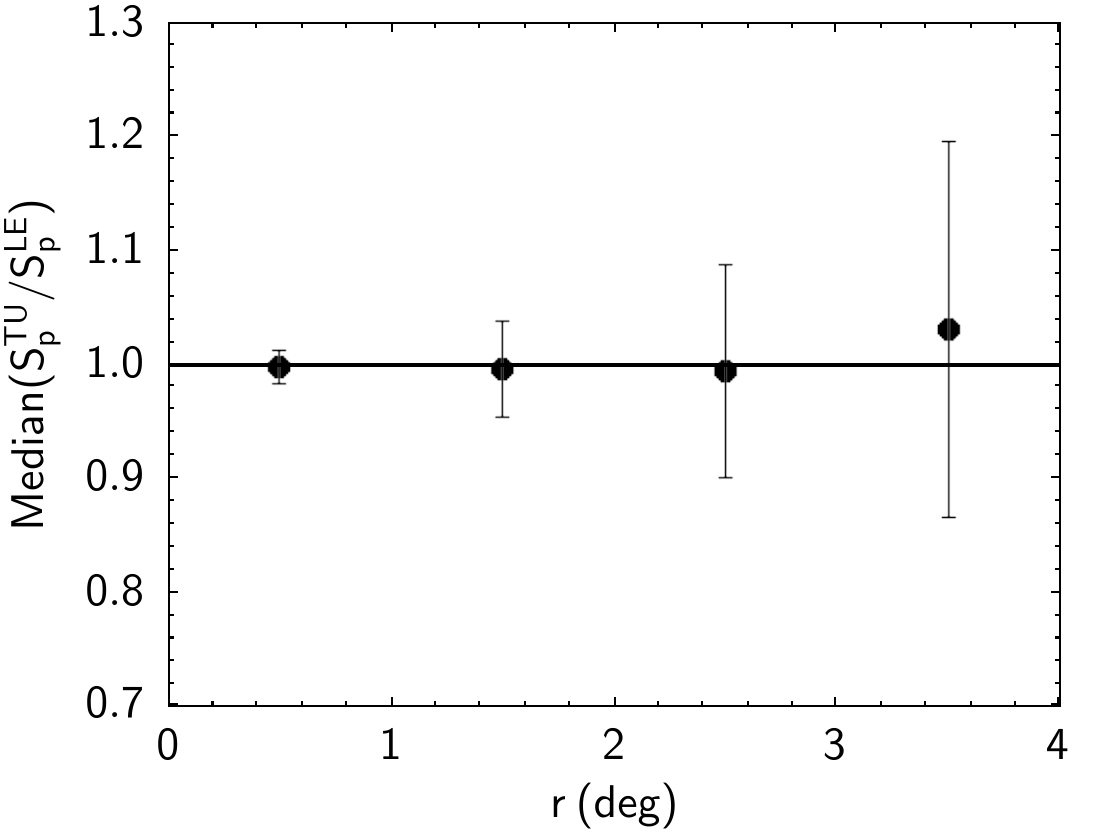}
 \includegraphics[width=6.0cm]{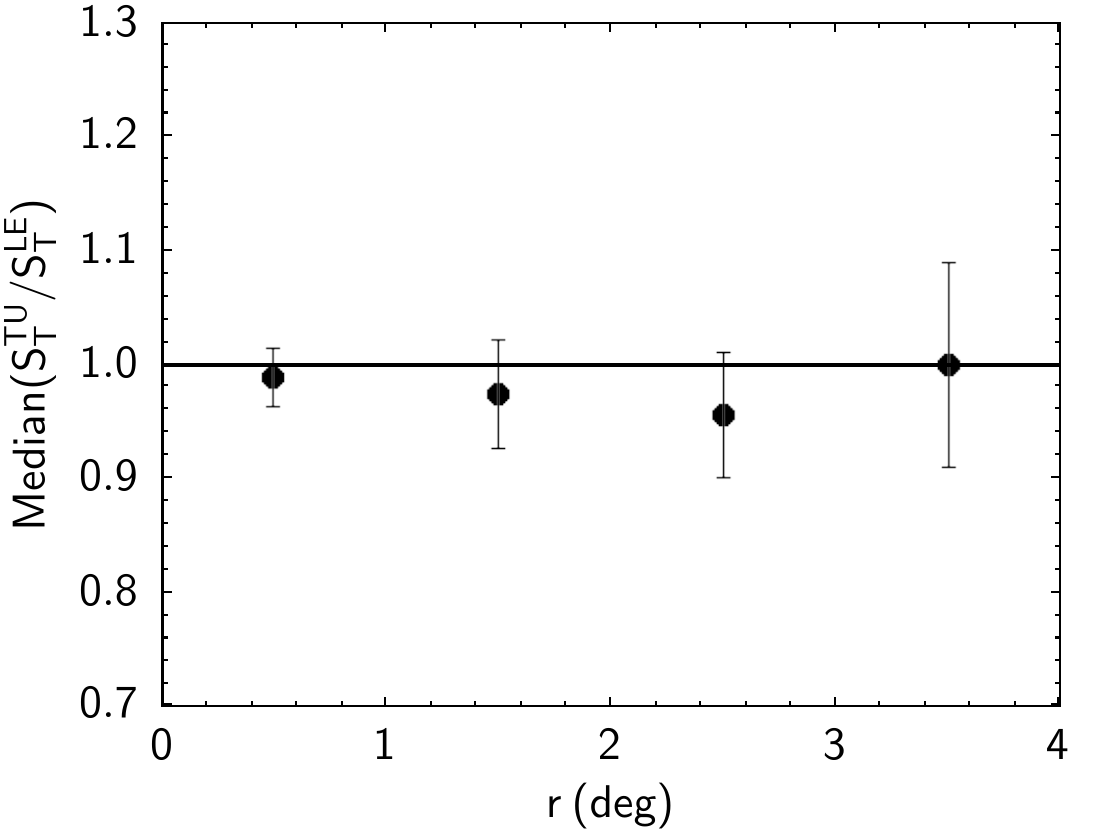}
 \includegraphics[width=6.0cm]{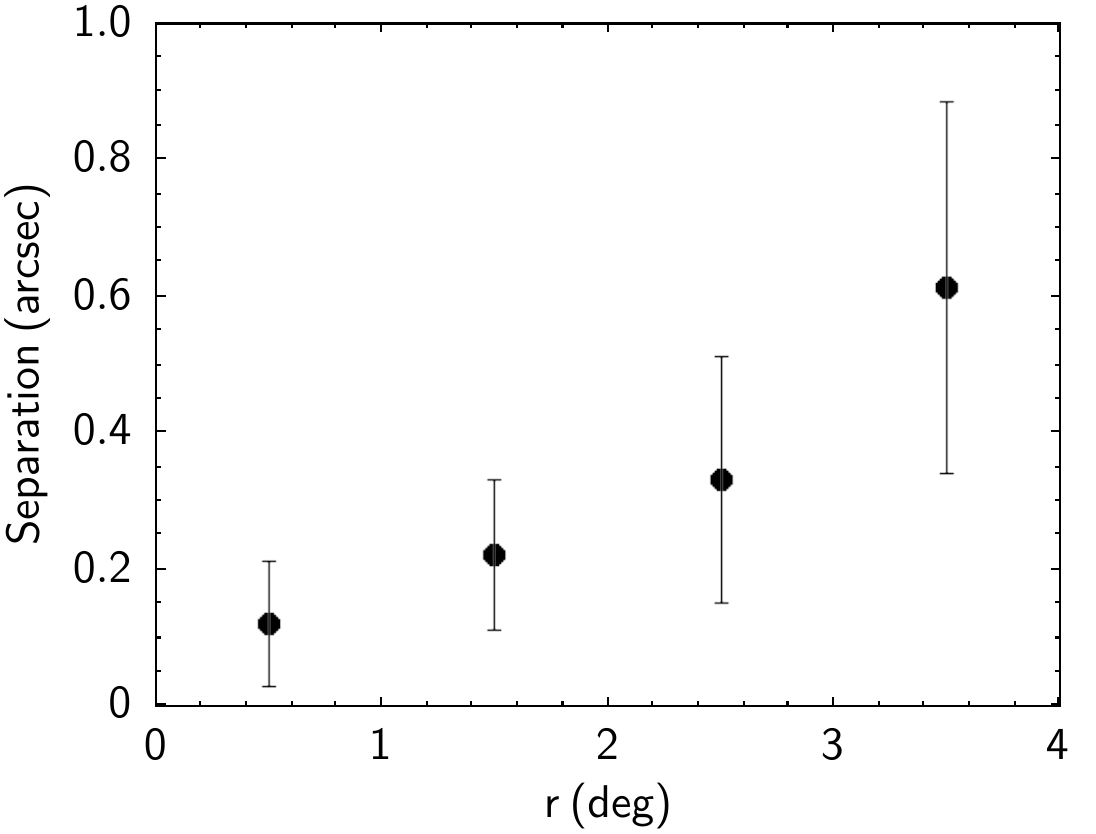}
\caption{Median peak brightness ratio (left panel), median total flux density ratio
 (middle panel) and median absolute value separation (right panel) in the four annular regions at increasing distance $r$ from the pointing centre for matched sources in the Turin and Leiden catalogues with SNR$>50$. The error bars are the scaled median absolute deviation (MAD) values.}
 \label{fig:ratio_flux}
\end{figure*}

The main results obtained from this comparison are shown in Fig.\ref{fig:ratio_flux} and can be summarised as follows: 
\begin{enumerate}
\item the median values of the
peak brightness ratios and total flux density ratios do not significantly change with increasing distance from the field centre and are consistent with a value of 1, meaning there is no
systematic scale offset in the amplitudes of the two images. 
\item
The MAD associated to each median value increases with $r$. Within $r<2$ deg the MAD is $<0.05$ for both the peak brightness ratios and total flux ratios. Beyond this distance the peak brightness ratios show a larger dispersion up to a MAD$=0.16$ in the range $3<r<4$ deg compared to a MAD$=0.09$ for the total flux density ratios in the same distance range. 
\item
We also checked whether the position of the radio sources in the Leiden and Turin catalogues could depend on $r$. 
The right panel in Fig.\ref{fig:ratio_flux} shows the median separation in arcsec as a function of distance $r$. We remind
the reader
that the two catalogues were matched using a maximum distance $r=3$ arcsec, half the size of the restoring beam. There
is a trend of larger differences in the source positions between the two catalogues with increasing distance from the field centre, but the median shifts are
small, $\lsim 0.2$ arcsec for $r<2$ deg and even at the largest distance probed by this test the median shift is $\sim 0.6$ arcsec, less than half of the pixel size.
\end{enumerate}
Summarising, the peak brightness and flux density ratios are consistent to 
$\leq 5\%$  within a radius $r=2$ deg from the field centre. 
However, it should be noted that
at larger distances from the field centre these uncertainties increase significantly, reaching values of about $16\%$ and $9\%$ at distance $r\simeq 4$ deg for the peak brightness and total flux density, respectively. 
The two images have been obtained using the same pipeline but on different computer infrastructures, slightly different singularity images, different starting sky models and faceting geometry. We note that we cannot exclude that the differences between the two images might be caused, at least to some extent, by different software versions or different computer hardware used to run the pipeline by us and in Leiden, but this is rather unlikely. The larger dispersion observed in the measured peak brightness ratios with respect to the total flux density ratios, and the 
general trend for the dispersion to be larger at larger distances
suggest that the differences between the two images
are likely caused by residual smearing and uncertainties in the direction-dependent calibration deriving from using different faceting patterns and starting sky models in the two images (see also the discussion in Sec.~\ref{sec:resunres}).

\subsection{Absolute flux density calibration}
\label{sec:scale}
We follow the method described in \citet{2021A&A...648A...2S} to adjust the absolute flux density scale using the external radio catalogues  available from the literature. We compared the flux density of sources in common between our catalogue and the external catalogues using only sources with distance $r<3$ deg from the field centre and adopting the constraints
listed in Section 3.5 from \citet{2021A&A...648A...2S}  to avoid the introduction of biases due to the different depths and angular resolution of the external catalogues. The external catalogues we used are the VLASSr at 74 MHz \citep{2014MNRAS.440..327L}, the TGSS at 150 MHz \citep{2017A&A...598A..78I}, the 6C at 151 MHz \citep{1990MNRAS.246..256H}, the WENSS at 350 MHz \citep{1997A&AS..124..259R}, the NVSS at 1.4 GHz \citep{1998AJ....115.1693C} and a WSRT pointed observation at 1.4 GHz \citep{2010A&A...517A..54W}.

\begin{figure}[htp]
 \centering
 \includegraphics[width=9.0cm]{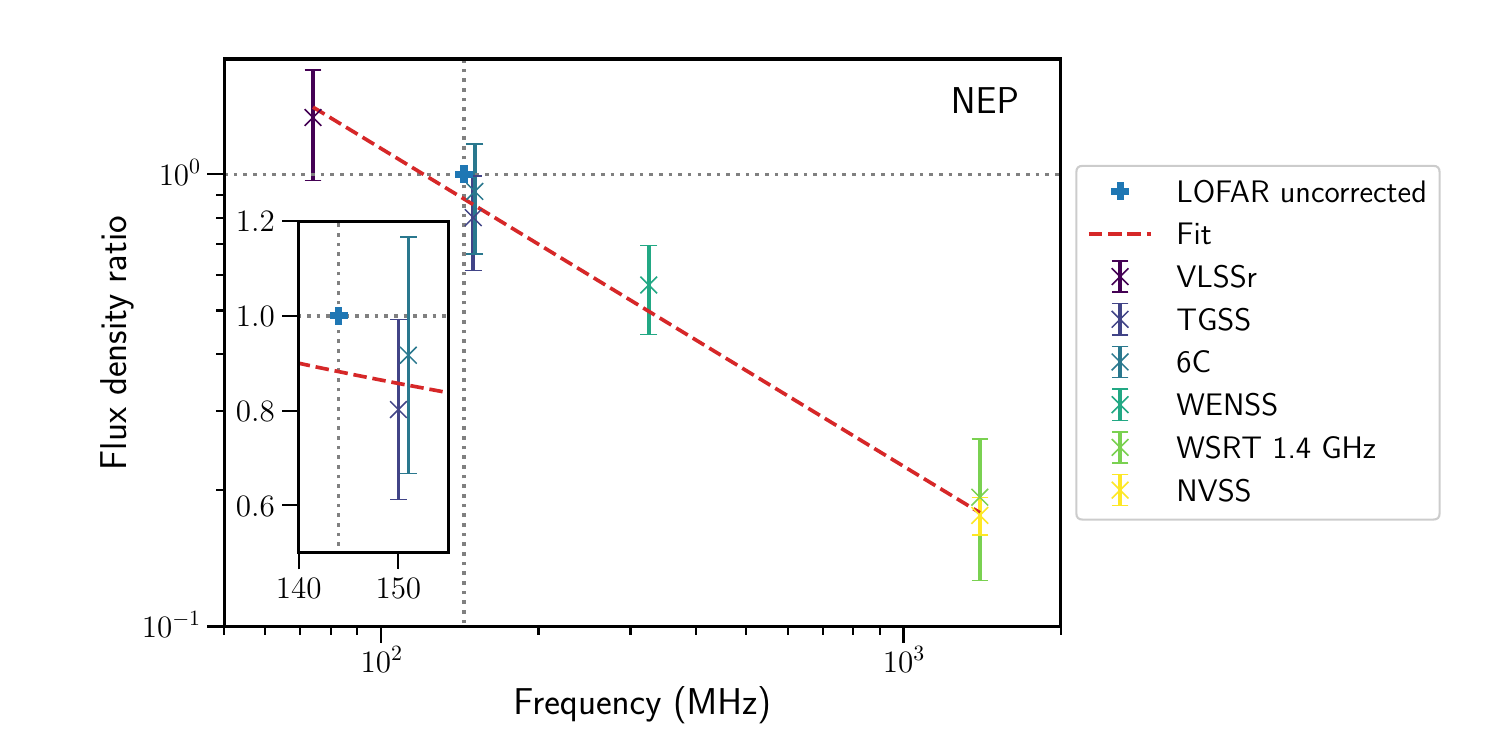}
\caption{Calibration of the EDFN flux density scale. The flux density scale after \texttt{\small DDF-pipeline} is set to unity and shown as a blue cross. The flux density ratios with respect to the external catalogues from the literature and their errors are shown in different colours. The red dashed line is the linear fit to the data. The inset shows a zoomed view close to the LOFAR HBA central frequency.}
\label{fig:flux_scale}
\end{figure}

The result is shown in Fig.\ref{fig:flux_scale}: each point is the median of the flux density ratios obtained from the LOFAR sources matched with each of the other external catalogues and the red dashed line is the linear best fit to these points. Assuming perfect a-priori calibration of the LOFAR array after the direction-dependent corrections, and of all other surveys plotted in Fig. \ref{fig:flux_scale}, and that a simple power-law spectral index is appropriate, the line should pass through the ratio value of 1 at 144 MHz. As can be seen from the inset in Fig.\ref{fig:flux_scale}, the best fit line has a value of $0.88\pm 0.04$. This is the scale factor that needs to be applied to the final image to correct the absolute amplitude scale.
The value we obtained for the EDFN is similar to those previously derived for other deep fields that are in the range 0.80-0.92 \citep{2021A&A...648A...2S}. We scaled the radio image by a factor of 0.88 to set the final LOFAR image of the EDFN to the correct amplitude scale. 

\subsection{The final catalogue of the Euclid Deep Field North}
\label{sec:catalogue}
Due to the intrinsic nature of the cleaning process, radio images can be contaminated by spurious sources
in the proximity of real bright radio objects. 
Furthermore, radio sources can have morphologies too complex and/or extended to be properly recovered by source finding algorithms which only provide a catalogue of radio components. 
Moving from a catalogue of radio components to a catalogue of radio sources requires some additional steps that will be discussed below.

Having the final radio image set to the correct flux scale we run again PyBDSF to obtain a new catalogue of radio sources, using the same parameters as before. Then we selected only the sources within a 10 deg$^2$ circular area  ($r=1.784$ deg)
centred on the Euclid Deep Field North position (RA=17:58:55.9 DEC=66:01:03.7), 
and throughout the rest of the paper we will consider only this area. 
This is the region that was originally selected
for the EDFN when the observations presented in this paper were planned, and only more recently the area was extended to 20 deg$^2$. It is worth saying that producing a catalogue from these observations covering the whole 20 deg$^2$ of the EDFN is not convenient, considering that the noise increases with the distance from the pointing centre and the effects discussed in Sec.~\ref{sec:ddeffects}. The more recent observations of the EDFN adopted a different pointing strategy that will increase sensitivity
and accuracy over a larger area allowing to properly investigate the full 20 deg$^2$ field.
In the final rescaled image the central r.m.s. noise
is $32\,\mu$Jy\,beam$^{-1}$, increasing to $\sim 45\,\mu$Jy\,beam$^{-1}$ at $r\simeq 1.8$\,deg due to the primary beam correction.

PyBDSF is a tool designed to decompose a radio image in islands and extract the components as a set of Gaussians, shapelets or wavelets, above a given threshold in SNR \citep{2015ascl.soft02007M}. Two or more radio components found inside the same island can be grouped together to form a single radio source on the basis of the distance between the components and the brightness distribution along the line joining the centre of the components. The outcome of this procedure is recorded in the output catalogue by the S\_Code parameter that is used to define the source structure: ``S\_Code=S'' for a single-Gaussian source that is the only component in the island , ``S\_Code=C'' for a  single-Gaussian source in an island with other sources, ``S\_Code=M'' for a multi-Gaussian source.
The raw catalogue produced by PyBDSF needs to be checked against known issues that can affect it and we briefly discuss them below, together with the solutions we adopted.
We supported our analysis by using the unWISE catalogue in the overlapping 10 deg$^2$ region \citep{2014AJ....147..108L,2017AJ....153...38M, 2017AJ....154..161M}.
The unWISE Catalogue is derived from the unWISE coadds of the Wide-field
Infrared Survey Explorer \citep[WISE,][]{2010AJ....140.1868W}
images, and includes two billion sources over the entire sky at 3.4 and 4.6 microns. We used the catalogue produced after 5 years of WISE imaging
\citep{2019ApJS..240...30S}. The unWISE catalogue has two advantages over the existing WISE catalogue (AllWISE): first, it is based on significantly deeper imaging, and second, it features improved modelling of crowded regions using the {\tt crowdsource}\footnote{ https://github.com/schlafly/crowdsource} analysis pipeline to simultaneously determine the positions and fluxes of all sources in the unWISE coadds.

\begin{figure*}[htp]
 \centering
 \includegraphics[width=6.0cm]{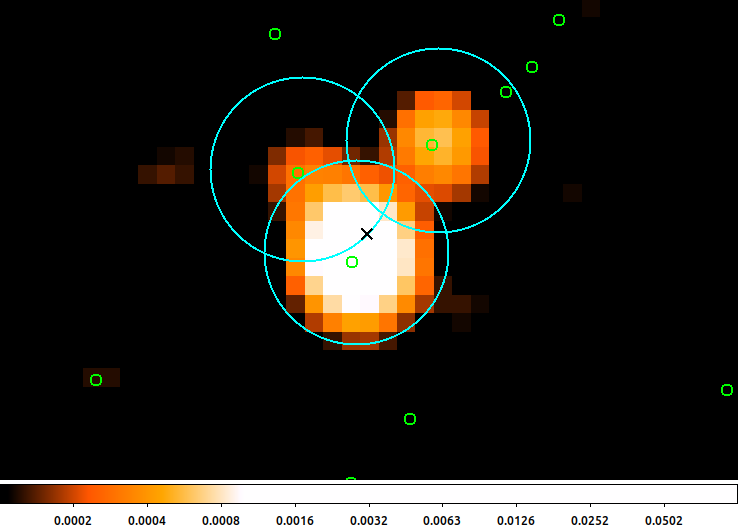}
 \includegraphics[width=6.0cm]{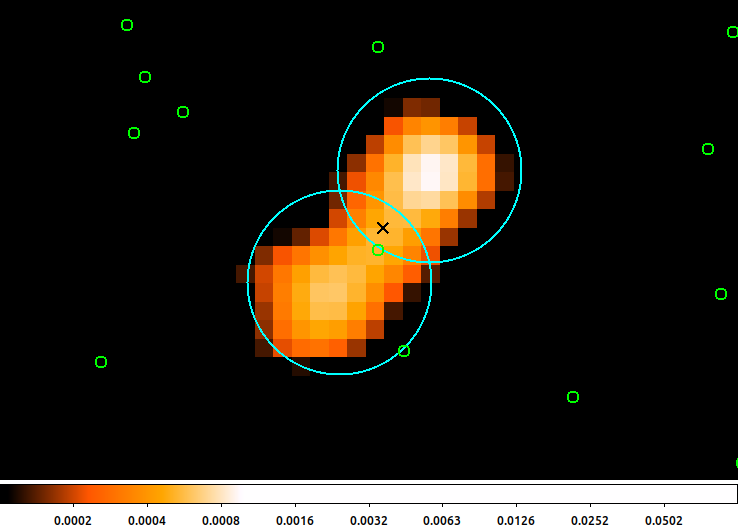}
 \includegraphics[width=6.0cm]{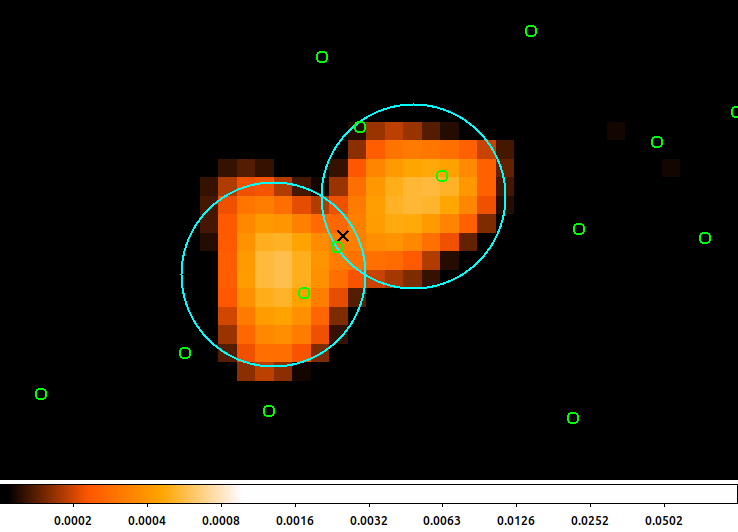}
\caption{LOFAR radio images at 144 MHz of three multi-component sources
each originally classified as a single radio source by PyBDSF. Cyan large circles identify the
individual Gaussian components, the black cross is the derived radio position
of the source, and the green circles mark the position of unWISE sources inside the region. The left panel shows a case where deblending is necessary as all the three components are very likely distinct radio sources. The middle panel is an example of correct association of different components to a single radio source, while the right panel is an uncertain case.}
 \label{fig:unwise}
\end{figure*}

Firstly, we check for spurious sources. Bright sources, in particular those where the direction dependent calibration did not perform well, can produce  artefacts associated with calibration uncertainties
that can exceed the local SNR threshold.
PyBDSF can generally deal with this issue adapting the window size over which to calculate the local r.m.s. around bright sources, but in some cases this is not effective.
These calibration artefacts produce a typical pattern with spikes irradiating from the bright source. We visually checked the region within a radius of $90$ arcsec around the sources with total flux $S_T> 15$ mJy looking for sources with $5< {\rm SNR}<10$ and without a counterpart in the unWISE catalogue (within a matching radius of $3$ arcsec). When such a source was found to be located along one of the spikes, it was considered a spurious source and it was removed from the catalogue. According to this criterion we removed 110 sources ($\sim 0.5$ percent).

As said above, when more than one Gaussian component is found within an island and a set of conditions are fulfilled, PyBDSF attempts to group together the components into a single radio source. This process cannot be perfect. For instance, two or more distinct radio sources can be grouped into a single multi-component source due to the limited resolution of the LOFAR observations. 
For this reason we visually inspected all the catalogued radio sources classified as ``M'' or ``C'' by PyBDSF, again using the unWISE catalogue to support our analysis.
Fig.~\ref{fig:unwise} shows the three typical cases we encountered. In each panel, the radio image is shown in colours, the cyan circles are centred at the positions of the different radio Gaussian components, the black cross is the final position
of the radio source derived from a brightness weighted average of the position of the individual components, and the green small circles indicate the positions of the unWISE sources in the area. In the left panel, the three Gaussian components are
grouped to form a single radio source by PyBDSF, but the comparison of the radio and unWISE source positions strongly suggests that we have three different radio sources. There were around 550 such cases where sources originally classified as a single source had to be split in two or more individual sources, in these cases we replace the single source in the radio catalogue with the individual Gaussian components that have  SNR$>5$. Deblended sources with SNR$<5$ were not included in the radio catalogue.  
In the example shown in Fig.~\ref{fig:unwise} all the three Gaussian components have SNR $>5$ and therefore the single entry is replaced by three entries with appropriate parameters in the final radio catalogue. The middle panel shows an example where there is an unWISE
source close to the radio source position and no unWISE sources near the peak of the individual components: in all these cases we considered the PyBDSF outcome as reliable. In the right panel, unWISE sources are found both at the radio source position and close to the peak of the individual components: such cases are uncertain and we decided not to change the PyBDSF output.

Another typical case of PyBDSF failure, in combining the appropriate components into a single radio source, is when components that belong to the same radio source are catalogued as distinct sources.
This can happen when the lobes of a very extended radio galaxy are separated
by tens of arcseconds or even arcminutes. Each of the two lobes (and a radio core, if present) can be classified as a distinct  source by PyBDSF, with each lobe usually classified as a multi-component source.  We visually inspected all the sources classified with ``S\_Code=M'' or ``S\_Code=C'', and we identified the lobe components of extended radio galaxies. Whenever necessary we 
grouped together the lobe components into a single source
in the catalogue.
We also searched for a possible radio core previously identified as a separate source and  we assigned it to the extended radio source. The total flux of the source was assumed to be the sum of the fluxes of all grouped components, and its position the one of the radio core. When no radio core is detected,
 the radio position is set equal to the
flux density weighted average of the position of the lobes.
For these 131 sources ($\sim 0.6\%$ of the total) the size parameters in the catalogue are set to -1 and the error associated with the total flux density is derived combining the errors of the individual components.
As a result of this process we obtain a final catalogue \footnote{The data associated with this article are released at:\\http:/lofar-surveys.org} over the 10 deg$^2$
field centred on the EDFN containing 23,333 sources ($93\%$ classified as ``S\_Code=S'') with SNR$>5$.

It is worth mentioning that the errors given in the catalogue are only those derived by PyBDSF from the fitting procedure. In particular, as far as concerns the peak brightness and total flux density values, these errors do not include the effects investigated in Sec.~\ref{sec:ddeffects} and Sec.~\ref{sec:scale}. Combining the error associated with the linear fit in Sec.~\ref{sec:scale} with the error on peak brightness or total flux density for sources within 2 degrees from the field centre we have a likely uncertainty of around 6 percent in the flux density scale, a value consistent with that
quoted for other deep fields \citep{2021A&A...648A...2S}.
\begin{figure}[htp]
 \centering
 \includegraphics[width=7.5cm]{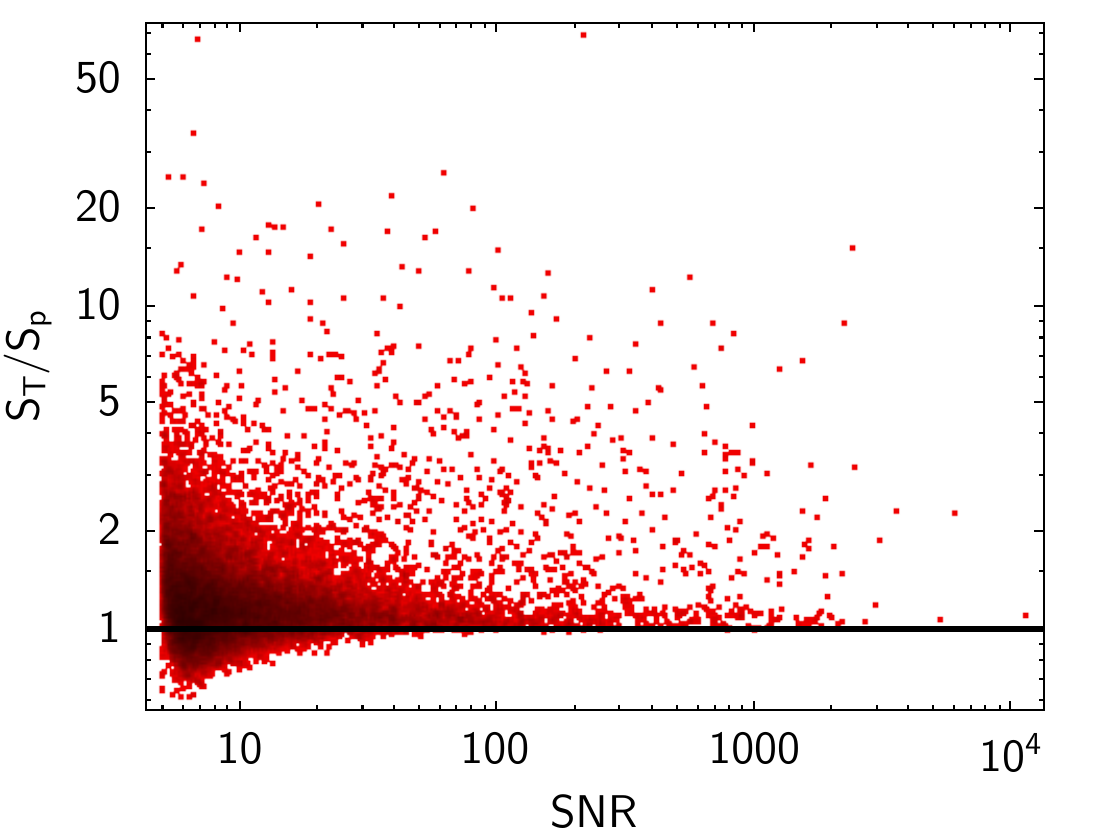}
 \includegraphics[width=7.5cm]{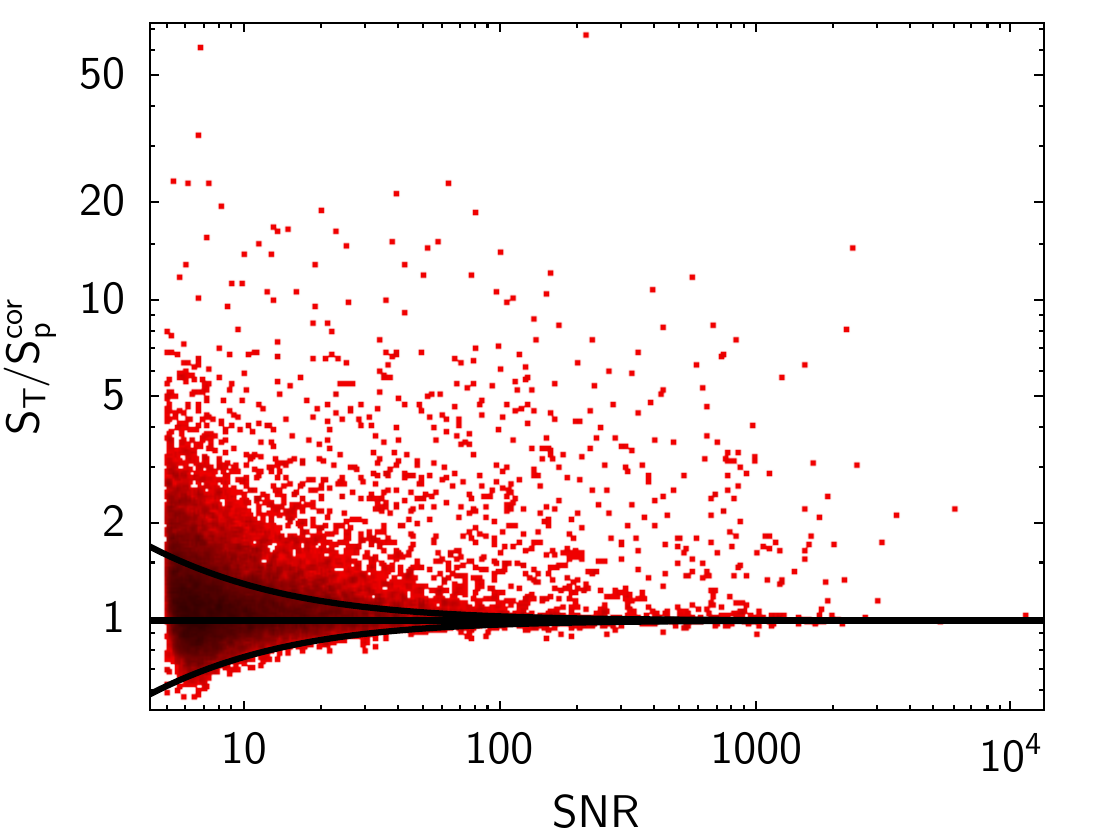}
\caption{The upper panel shows the total-to-peak flux ratio vs signal-to-noise ratio plot for all the sources in the catalogue. The
line $S_T/S_p = 1$ is drawn.
The lower panel shows the same plot using $S^{\rm cor}_p$ instead of $S_p$. $S^{\rm cor}_p$ is the corrected peak flux obtained by multiplying $S_p$ by a radially dependent correction factor
(see Sec.~\ref{sec:resunres} for details).
The upper black line separates resolved from unresolved sources according to the relation $S_T/S^{\rm cor}_p > (1+3/{\rm SNR})$. The lower black line shows the mirrored relation.}
\label{fig:resunres}
\end{figure}
\begin{figure}[htp]
 \centering
\includegraphics[width=5.7cm]{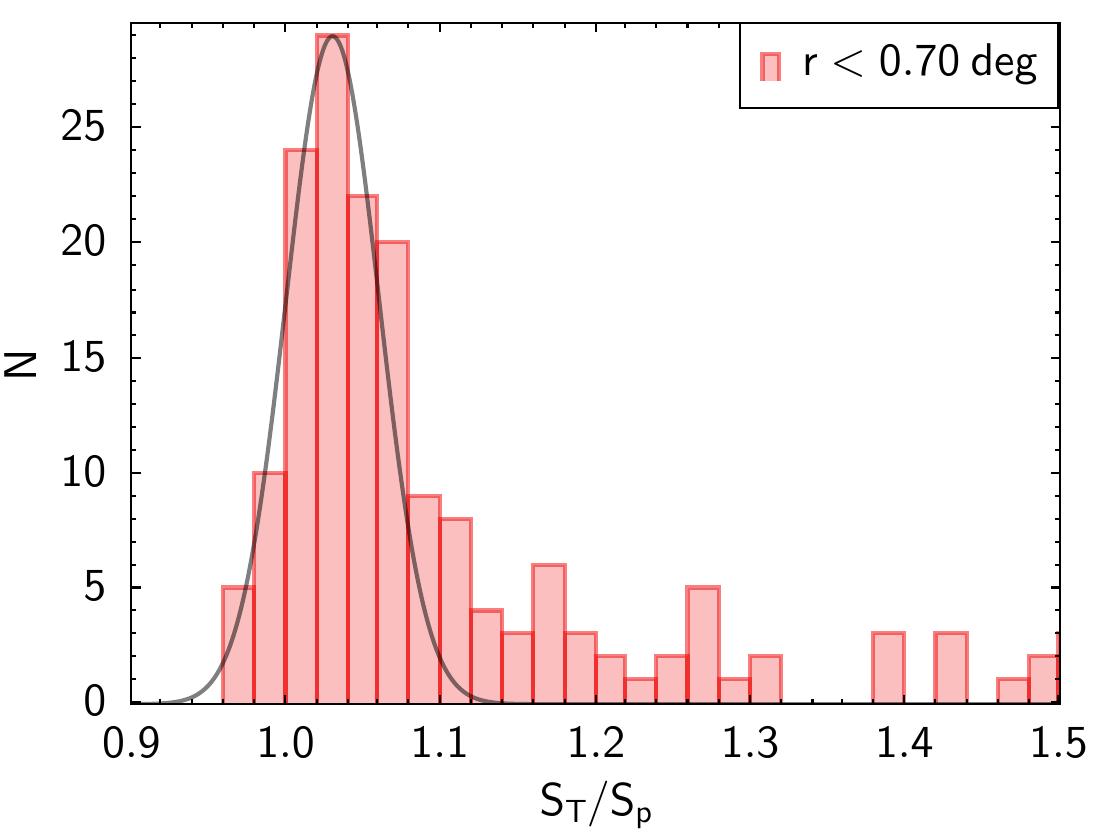}
\includegraphics[width=5.7cm]{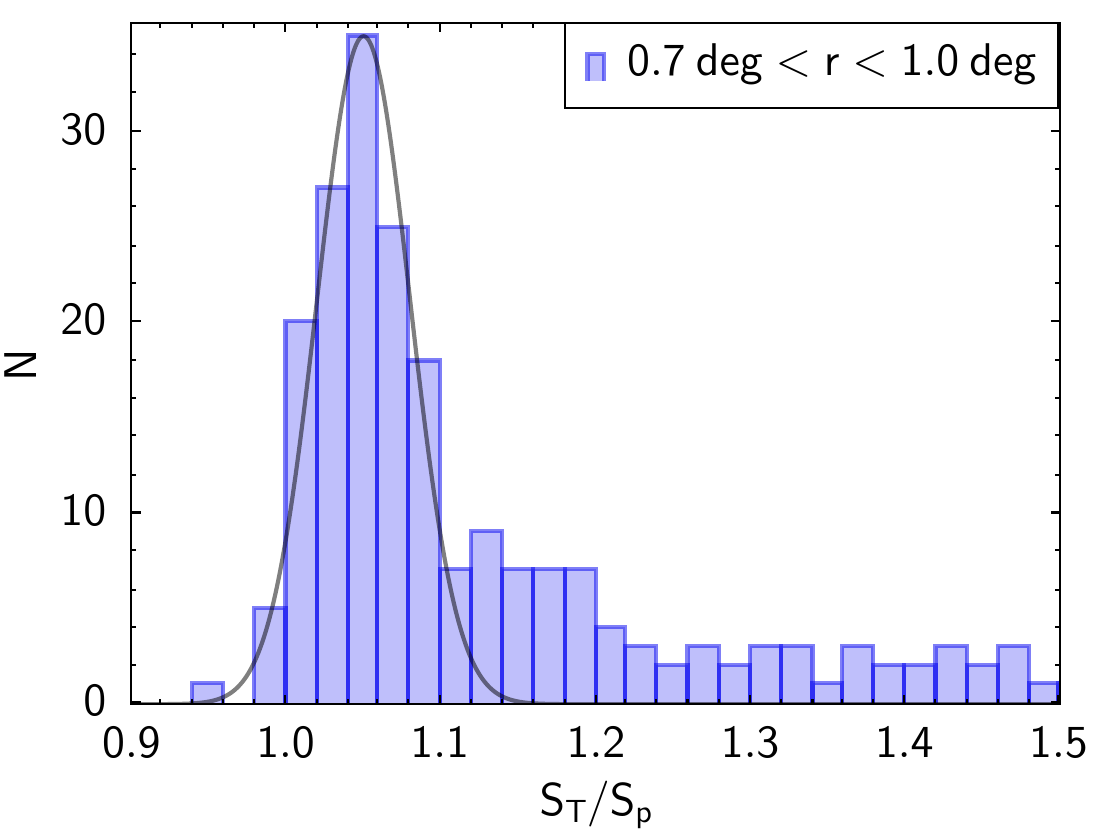}
\includegraphics[width=5.7cm]{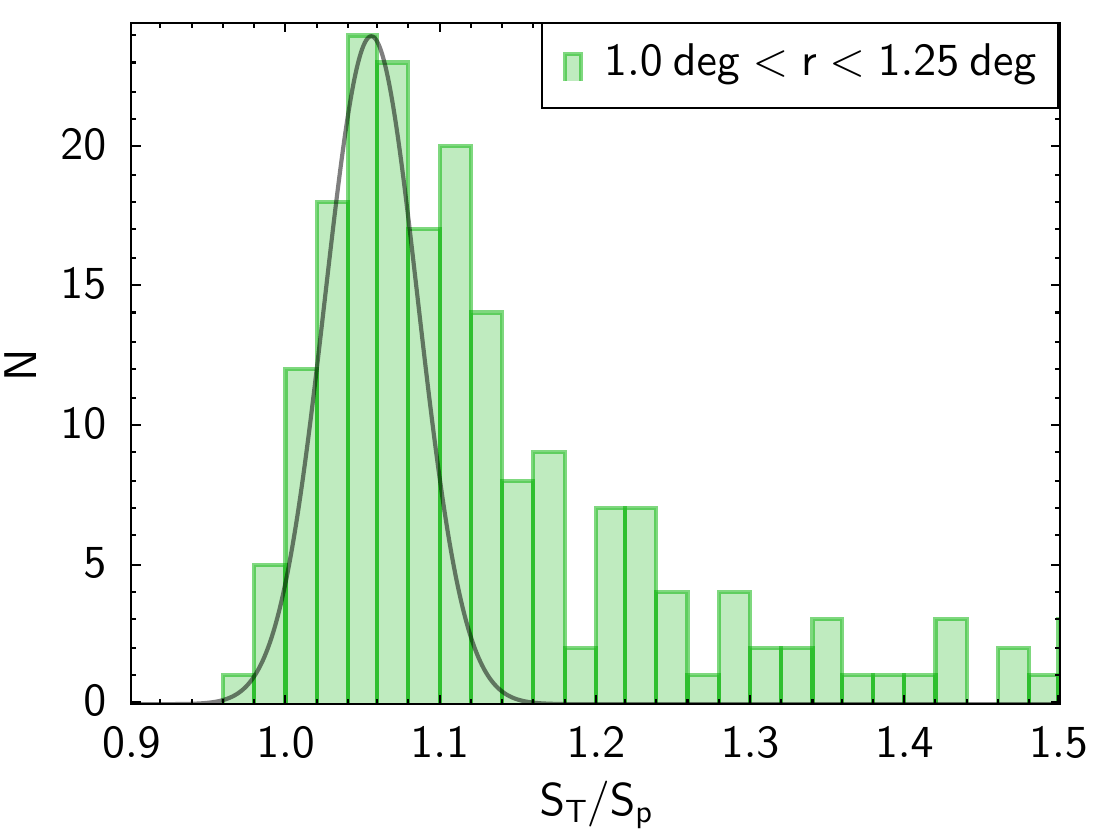}
\includegraphics[width=5.7cm]{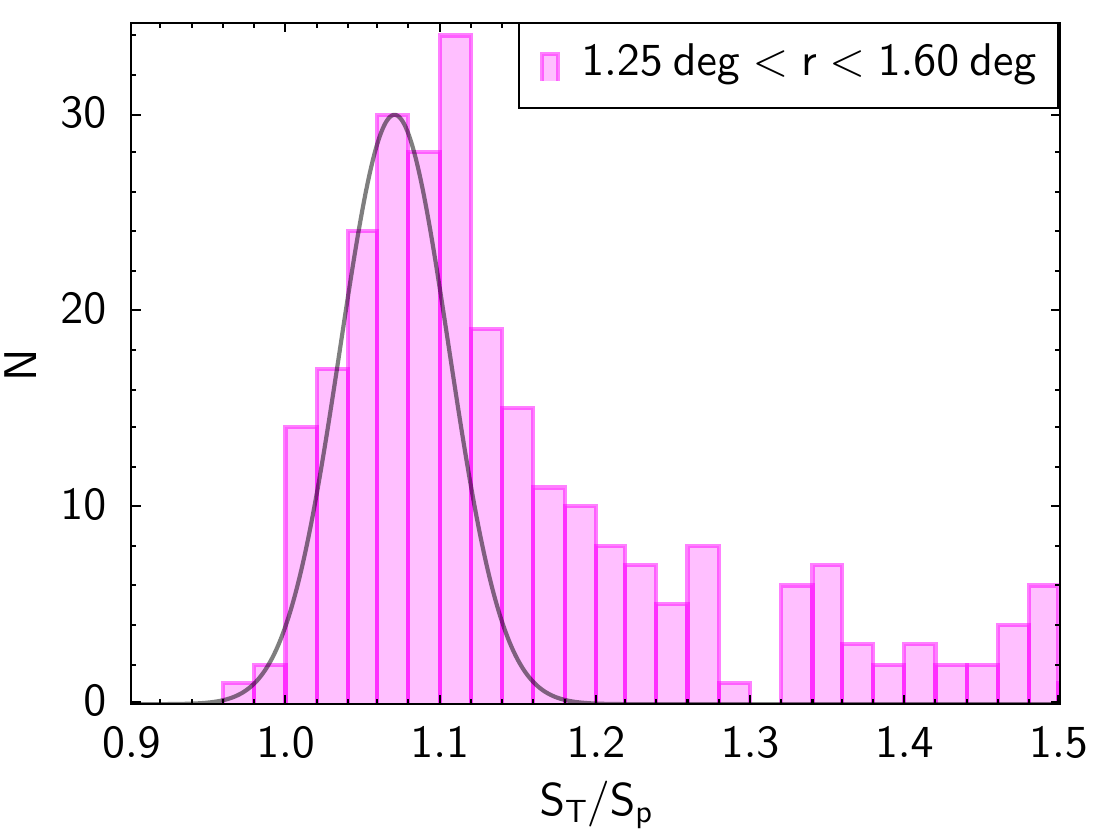}
\includegraphics[width=5.7cm]{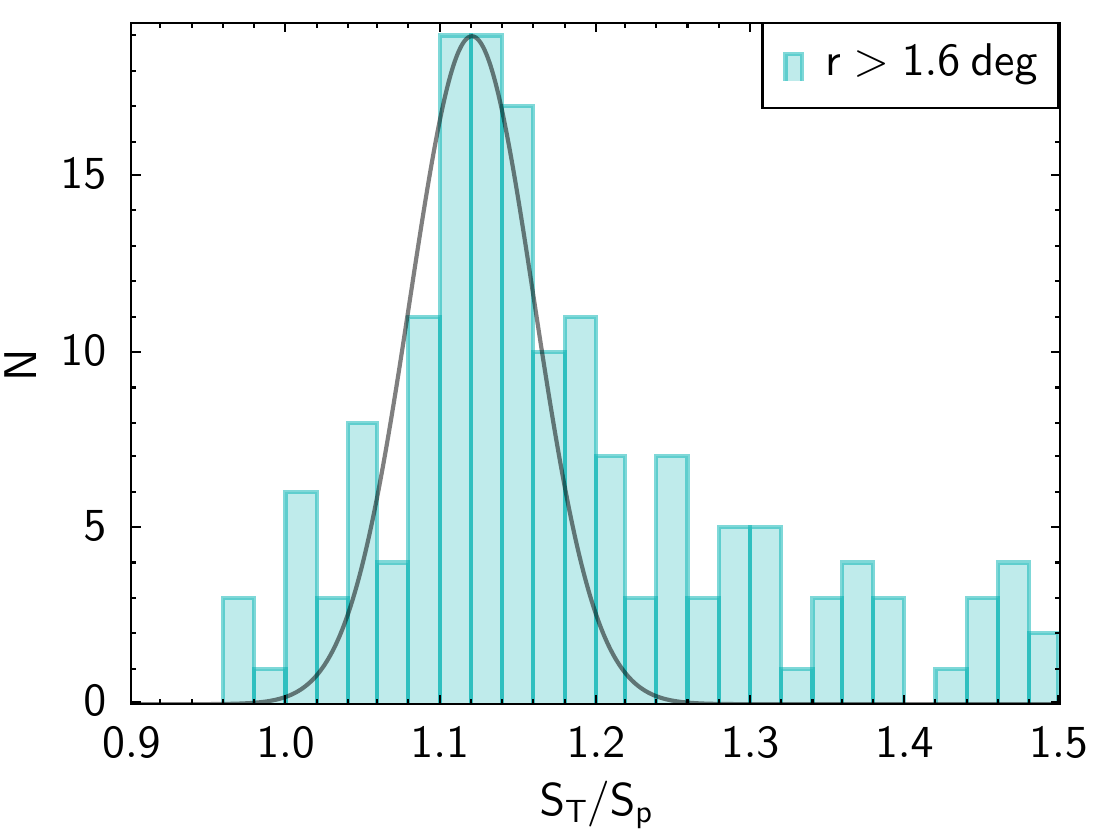}
\caption{Distribution of $S_T/S_p$ values for sources in different bins of distance from the field centre. Only sources with SNR$>40$ are considered. The plots are truncated at $S_T/S_p=1.5$ for a better visualization. For each distance bin we fit a Gaussian function to the region of low $S_T/S_p$ values.} 
\label{fig:resunres2}
\end{figure}


\subsection{Resolved and unresolved sources}
\label{sec:resunres}
The ratio between measured total and peak fluxes ($S_T/S_p$) is a proxy for the extension of a radio source since it is equivalent to the ratio between the fitted source full width half maximum (FWHM) axes, $\theta_{\rm maj}$ and $\theta_{min}$, and the restoring beam FWHM axes, $b_{\rm maj}$ and $b_{\rm min}$:
\begin{equation}
 S_T/S_p = \theta_{\rm maj}\theta_{min}/b_{\rm maj}b_{\rm min}
\end{equation} 
In Fig.~\ref{fig:resunres} (upper panel) we plot  $S_T/S_p$ versus the SNR for all the sources in the final catalogue.
Values of $S_T/S_p <1$ are due to 
statistical errors affecting the flux density measurements and these errors should equally affect the $S_T/S_p >1$ region. 
In the following we use $S_T/S_p$ to classify the sources as resolved or unresolved.
It is also clear that high SNR sources systematically lay above the $S_T/S_p=1$ line (shown in black in Fig.~\ref{fig:resunres}). 
This is a clear indication that the distribution of $S_T/S_p$ is affected by an offset that artificially increases the derived ratios. Such an offset has been previously found, for instance, in measurements derived from images affected by bandwidth smearing \citep[e.g.][]{2008ApJ...681.1129B} and in the
LOFAR observations of the Lockman Hole and ELAIS-N1, where it has been interpreted as due to a combination of residual facet-dependent calibration errors and PSF modelling \citep{2021A&A...648A...5M}.

\citet{2021A&A...648A...5M} accounted for the statistical errors in $S_T/S_p$ by defining a lower envelope of the $S_T/S_p$ distribution shown in Fig.~\ref{fig:resunres}, as:

\begin{equation}
 S_T/S_p = A/(1+B/{\rm SNR})
\end{equation}
where the parameter A represents the aforementioned offset.
The lower envelope can then be mirrored around the $S_T/S_p=A$ axis to get the upper envelope:

\begin{equation}
 S_T/S_p = A\times(1+B/{\rm SNR})
\end{equation}
All the sources above the upper envelope are classified as resolved,
while those below are considered unresolved. 
\citet{2021A&A...648A...5M} derived single A and B values for each of the LoTSS Deep Fields.

We decided to use a slightly different approach.
To investigate and quantify the offset A we split all the sources with SNR$>40$ in the final catalogue according to their distance from the field centre ($r$) into 5 different subsamples: $r<0.7$ deg, $0.7 < r < 1.0$ deg, $1.0<r<1.25$ deg, $1.25<r<1.6$ deg, and $r>1.6$ deg. 
For each subsample we plot the distribution of the ratio $S_T/S_p$ in Fig.~\ref{fig:resunres2}. The plots are truncated at $S_T/S_p =1.5$, but the tail of extended sources continues up to
value of $\sim 50$. Each distribution shows a Gaussian-like region at low $S_T/S_p$ values that we fit with a Gaussian (the black curves in Fig.~\ref{fig:resunres2}).

We found that the Gaussian region of the $S_T/S_p$ distribution has 
an offset with respect to the $S_T/S_p =1$ value and that this offset changes for sources at different distances from the field centre. More distant sources show a higher value (A $\sim 1.12$) than the sources closer to the field centre (A $\sim 1.03$).
In Tab.~\ref{tab:offset} we list for each subsample the range in $r$, the number of sources in that distance range, and the A values derived from the Gaussian region of the $S_T/S_p$ distributions.

The radio catalogue lists sources in a 10 deg$^2$ circular region ($r < 1.784$ deg) and, given the spectral (97.6 kHz) and time (8s) averaging of the data, we can expect bandwidth and time smearing up to $20\%$ at the edges of the field and around $8\%$
at 1 deg from the field centre \citep{1999ASPC..180..371B}.
\texttt{\small DDFacet}, the imaging code within \texttt{\small DDF-pipeline}
is specifically designed to minimize the decorrelation and  smearing effects using facet-dependent corrections during the deconvolution \citep{2018A&A...611A..87T}.
Such a radially dependent residual smearing was already noted by
\citet[see their Fig.10]{2019A&A...622A...1S} in the LoTSS images.
\texttt{\small DDFacet} derives, for each facet, its own PSF to be used during deconvolution, but large
facets (spanning a significant range in $r$) can still produce 
smearing at the levels we observe. Moreover, it is worth noting that facets at larger distances tend to be larger. 
As we have shown in  Sec.~\ref{sec:ddeffects}, the accuracy of the flux measurements decreases with increasing distance from the field centre, and this effect is stronger for peak fluxes compared to total fluxes as expected in case of smearing.
\citet{2021A&A...648A...5M} arrived at the same conclusion, with the only difference that they used a single A value, averaged over distance.
We conclude that the observed total-to-peak flux ratio offset is caused by a residual smearing effect, which is not totally corrected by {\tt DDFacet}. 

 In order to separate resolved from unresolved sources, we first correct the source peak fluxes in the final catalogue using the A values listed in Tab.~\ref{tab:offset} according to the distance of the source from the field centre. The values in Tab.~\ref{tab:offset} are reasonably well fitted by a parabola given by $A(r)=1.025+(r^2/35)$, where $r$ is the distance from the field centre in degree, and we use this expression to interpolate the corrected peak fluxes ($S^{\rm cor}_p$) at each distance. The lower panel in Fig.~\ref{fig:resunres} shows the new $S_T/S^{\rm cor}_p$ versus SNR plot. The $S_T/S^{\rm cor}_p$ distribution, for the sources at high SNRs, has a Gaussian region centred at $S_T/S^{\rm cor}_p=1$. Then we assume for the lower envelope: 

\begin{equation}
 S_T/S^{\rm cor}_p = 1/(1+3.0/{\rm SNR})
 \label{eq:resunres_low}
 \end{equation}
where $B=3.0$ was chosen to be comparable with the  equivalent functions in \citet{2021A&A...648A...5M} and satisfying the criterion that only $\simeq 1$ percent of the points are outliers (i.e.
sources below the lower envelope).
Mirroring Eq.~\ref{eq:resunres_low} we obtain the selection function for the resolved sources:
\begin{equation}
 S_T/S^{\rm cor}_p = 1+3.0/{\rm SNR}
 \label{eq:resunres_high}
 \end{equation}

Adopting this relation we obtain that $\simeq 24\%$ of the sources are classified as resolved. This value is consistent with the values of resolved sources
derived in other LOFAR deep fields as Lockman Hole and Elais-N1 \citep{2021A&A...648A...5M}, and with the fraction of $24\%$ of resolved sources found in the VLA-VIRMOS VLT Deep Survey \citep{2003A&A...403..857B}, a survey made with the VLA at 1.4 GHz, but with the same angular resolution (6 arcsec) of LOFAR HBA images.
For the purposes of deriving the radio number counts (see Sec.~\ref{sec:counts}) we will use the total flux density ($S_T$) for the $\sim 5,500$ resolved sources and the corrected peak flux ($S^{\rm cor}_p$) for the  $\sim 17,800$ unresolved sources. Needless to say that the method adopted to classify a source as resolved or unresolved is an approximation. One limitation is that we assume that all the sources below the upper envelope of Fig.~\ref{fig:resunres} but with $S_{\rm T}/S^{\rm cor}_p>1$ are unresolved and this is likely not be entirely true. Another one is how to fit the lower envelope.  For all these reasons the
classification as resolved or unresolved is used only on a statistical basis and the total flux density derived from the classification is used only to calculate the source counts and not listed in the catalogue. The fluxes in the catalogue are those derived by fitting and assembling Gaussian
components by PyBDSF or manually for the very extended sources. 
Having said that, we note that using samples of mock sources to derive the source counts corrections allows to
correct, if not entirely, at least for part of these  limitations as we explain in the next section. 
\begin{table}
\caption{$S_T/S_p$ offset values}
\centering
\begin{tabular}{rcc}
\hline
Distance interval& Number& A($r$) \\
(deg)            &       &    \\ 
\hline
$r<0.70$          & 232 & 1.030 \\
$0.70 < r < 1.00$ & 270 & 1.050 \\
$1.00 < r < 1.25$ & 266 & 1.055 \\
$1.25 < r < 1.60$ & 394 & 1.070 \\
$r>1.60$          & 230 & 1.120 \\
\hline
\end{tabular}
\tablefoot{Col.1: Distance from the field centre bin; Col.2: number of SNR$>40$ sources in the distance bin; Col.3: Offset value of $S_T/S_p$ derived from the peak of the Gaussian fit.}
\label{tab:offset}
\end{table}

\section{Radio source counts corrections}
The observed number counts of radio sources must be corrected for several
effects such as:
\begin{itemize}
 \item 
  varying noise distribution in the radio image due to the presence of bright sources and the effect of the primary beam correction.
 \item
 systematic effects in the source extraction procedure that could affect the measured peak brightness and total flux density.
 \item
 the resolution bias:
 number counts are derived in bins of total flux density from samples of radio sources that are selected on the basis of their peak brightness. For this reason the completeness of the radio sample depends also on the intrinsic angular size distribution.
 \item
 the Eddington bias \citep{1913MNRAS..73..359E}, also referred to as noise bias.
 \item
 the contribution of spurious sources.
\end{itemize}

In order to take into account the combined effects of the first four effects we used mock samples of radio sources as described below
(see Sec.~\ref{sec:simul}).
The contamination by spurious sources is not modelled by the simulations, since the mock samples are inserted in the same residual image obtained after subtraction of all the components.
The fraction of spurious sources is derived in the next subsection.

\begin{table}
\caption{False detection rates}
\centering
\begin{tabular}{rc}
\hline
SNR     & Fraction range \\
        &      (\%)       \\  
\hline
5.0--5.5 & 2.6--3.6 \\
5.5--6.0 & 1.2--1.8 \\
6.0--6.5 & 0.7--1.1 \\
6.5--7.0 & 0.4--1.1 \\
7.0--8.0 & 0.6--1.1 \\
8.0--10.0& 0.5--0.9 \\
10.0--20.0& 0.1--0.4 \\
\hline
\end{tabular}
\tablefoot{Col.1: SNR bin; Col.2: range of false detection rates, the upper value is obtained using all sources with
SNR$>5$ in the inverted image, while for the lower value only those without a bright source within $90\asec$ are used.}
\label{tab:false}
\end{table}

\subsection{False detection rate}
\label{sec:false}
The false detection rate is associated to spurious sources that are detected above the SNR threshold value.
To derive an estimate of the fraction of spurious sources still present in the catalogue we ran PyBDSF on the inverted (i.e. multiplied by 
$-1$) LOFAR image with the same settings used to produce the catalogue.
Since there is no negative emission in the radio continuum emission of the sky,
every detection above $5\sigma$ in the inverted image is due to a noise peak.
Therefore, assuming the noise distribution is symmetric around zero we expect a similar number of
noise peaks above $5\sigma$ (false detection rate) in the positive image as well.
This assumption may not be valid near the bright sources due to the combination of residual calibration errors and the shape of the sidelobes of the beam (i.e. the peak of the positive sidelobes is
higher than the absolute value of the peak of the negative sidelobes). Therefore, the number of negative peaks has to be considered as a lower limit for the number of spurious sources. For this reason in Sec.~\ref{sec:catalogue} we previously removed 
spurious radio sources with SNR$>5$  produced or contaminated by the secondary lobes of nearby bright objects ($S_T>$ 15 mJy). 
PyBDSF returned 236 negative peaks above  $5\sigma$ within the 10 deg$^2$ region. 
Among these 236 negative peaks, 101 are found within a distance of $90\asec$ from a bright source in the positive image. This number is close to the number of sources (110) that were excluded as spurious in Sec.~\ref{sec:catalogue}. We derive the fraction of false detections using
both the 135 sources far from bright sources (as a lower limit) and the whole sample of 236 sources.  The sources were binned in SNR groupings alongside true detections and the false detection rates, defined as the number of false detections divided by the number
of recovered sources in each SNR bin, are reported in Table~\ref{tab:false}. In general, the false detection rate values derived are rather small. We expect $\sim 3\%$
of false detection rate for sources in the range $5.0\le {\rm SNR} <5.5$ and about $\sim 1.5\%$ in the range $5.5\le {\rm SNR} <6.0$. For sources with SNR $>10$ the false detection rate is $<0.5\%$, so practically negligible.
We applied a correction due to the false detection rate to the
differential source counts listed in Table\,\ref{tab:counts} and Fig.\,\ref{fig:counts} using all the 236 false detections split into the appropriate total flux density bins. We note that the differences in the source counts deriving from using 135 or 236 false detections are much smaller than the error bars.

\subsection{Completeness and noise bias corrections}
\label{sec:simul}

To derive the completeness and noise bias corrections affecting the EDFN radio catalogue we follow the same approach used, for instance,  for the COSMOS-3GHz catalogue \citep{2017A&A...602A...1S}.
This procedure requires generating a realistic sample of mock radio sources that will be added to a radio image with the same noise properties of the real one. Then, the new image containing the mock sources is processed with the same algorithm that yielded the radio source catalogue. Sources in the input and recovered catalogues are then split into bins of total flux density, and the numbers of injected versus recovered sources in each total flux density bin are compared.
The image on which the mock sources are injected is the residual image produced by PyBDSF after the real sources have been extracted. 
To limit the effects of confusion and blending of different radio sources (the effect we, partially, correct for using the UnWISE images as described in Sec.~\ref{sec:catalogue} for the real sources) we set a minimum distance of 12 arcsec between two mock sources.
With the expression realistic
sample we mean a sample of mock sources that follows a flux density and angular size distribution as close as possible to those
of real radio sources as described in the next subsections.

\subsubsection{Mock radio sources catalogue: flux density distribution}
To simulate the flux density distribution we used  a 7-th order polynomial function fitting the differential source counts derived from the other LoTSS Deep Fields and the TGSS-ADR1 
\citep{2017A&A...598A..78I} to better constrain the bright end of the counts
\citep[eq. 13 and Table 4 in ][]{2021A&A...648A...5M}. The mock
catalogue is generated down to a flux density of $90\,\mu$Jy (roughly 
$3\sigma$) well below the $5\sigma$ detection threshold used to produce our radio catalogue. We generated three different realisations of the mock source catalogue that were individually processed and compared. Each mock catalogue contains $\simeq 55,000$ sources in the 10 deg$^2$ area.

\subsubsection{Mock radio sources catalogue: angular size distribution}
The most challenging aspect in generating mock samples of realistic radio sources is to assign to each source its own angular size. This is because the intrinsic source angular size distribution of sub-mJy radio sources is still not well known. The modelling of the intrinsic angular size distribution of the radio sources is necessary to correct for the resolution bias which can severely affect the radio source counts, since these are a function of the total flux density, while the completeness of a radio catalogue is typically based on the signal-to-noise ratio and, therefore, on the peak brightness. Clearly, such an effect is more severe for catalogues derived from observations with an angular resolution $\theta\lsim 1\asec$, but it needs to be taken into account in the lower resolution ($\theta\simeq 6\asec$) LOFAR observations as well. For a more detailed
description of the methods used to overcome this issue we refer to
\citet{2008ApJ...681.1129B} and \citet{2017A&A...602A...1S}.

To allow for an easier comparison with the results obtained in the other LoTSS Deep Fields we adopted  the radio source angular size distribution modelled by eqs (7) and (9) in \citet{2021A&A...648A...5M} with $m=0.3$.
Using this distribution we assigned to each source its own angular size and we modelled the mock sources as circular Gaussian. It is worth noting that real radio sources usually are not circular Gaussians. 
The effects on the radio source counts produced by a parent source population with more realistic source morphologies have been recently investigated by \citet{2023MNRAS.520.2668H}. Using rather low resolution ($\simeq 8\asec$) images, \citet{2023MNRAS.520.2668H} find that the source counts are not strongly affected by different source models. This is not surprising given the low resolution, that is comparable to that of our LOFAR image. At these resolutions the bulk of the radio sources detected in a deep field are unresolved or slightly resolved
and the circular Gaussian approximation  is appropriate.
Such an approximation is not valid anymore for catalogues of radio sources derived from images at higher resolution and in such a case a more detailed source modelling is necessary \citep[e.g.][]{2017A&A...602A...1S}.
\begin{figure}[htp]
 \centering
 \includegraphics[width=7.3cm]{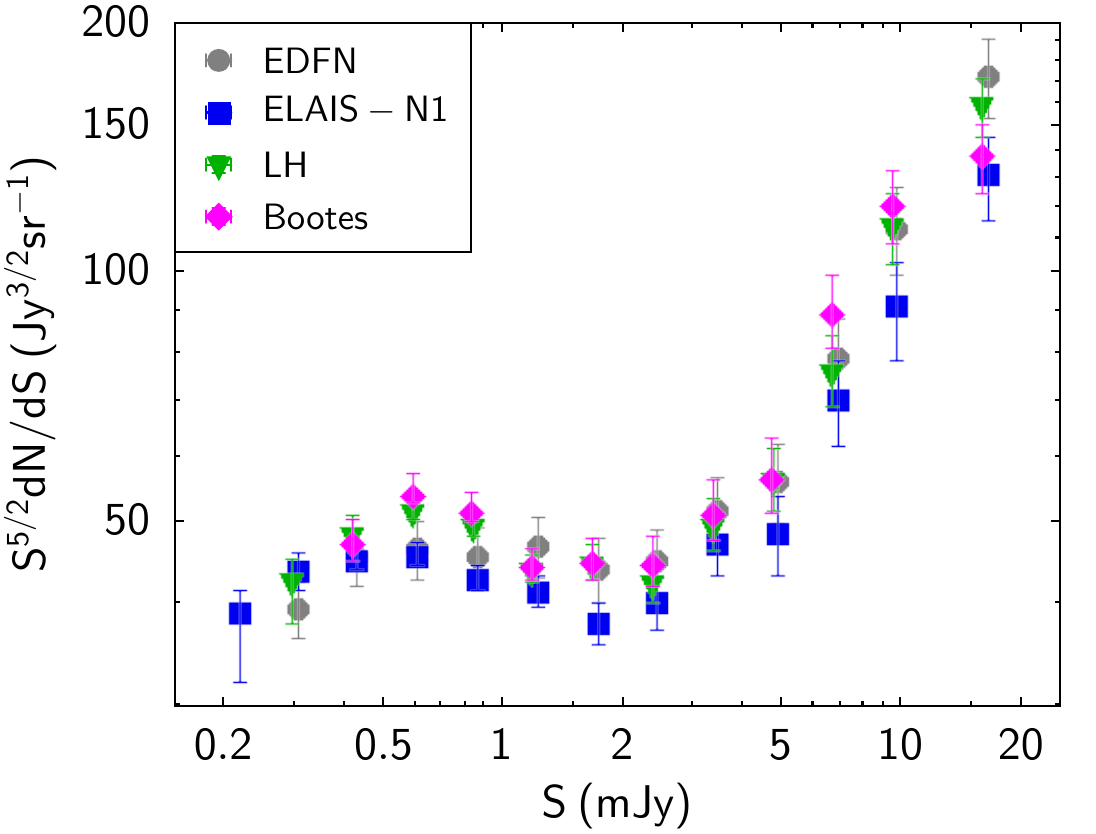}
 \includegraphics[width=7.3cm]{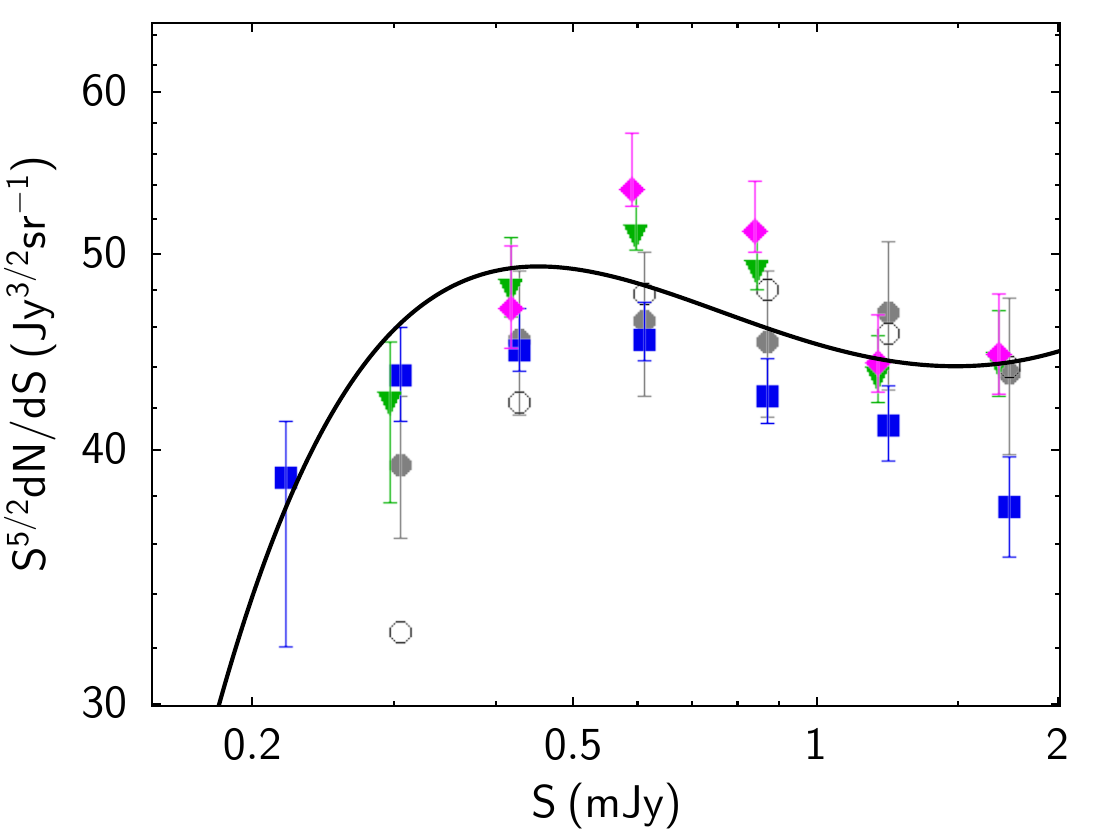}
\caption{Euclidean normalized radio source counts at 144 MHz, obtained for the Euclid Deep Field North plotted together with those obtained from other LOFAR Deep Fields (see symbols) from \citet{2021A&A...648A...5M}. The black line is the 7-th order polynomial fit from \citet{2021A&A...648A...5M}. The lower panel is a zoom of the region between 0.2 and 2 mJy showing also with empty circles the uncorrected normalized source counts for the EDFN.}
\label{fig:counts}
\end{figure}
\begin{table*}
\caption{Differential source counts in the EDFN}
\centering
\begin{tabular}{rrrrrrrrr}
\hline
S$_{\rm min}$ & S$_{\rm max}$&$\Delta$S&S&N$_S$&$n$&$n_{\rm cor}$& err$_{\rm y}$& $C_{\rm sim}$ \\
\hline
0.26 & 0.36 & 0.10 & 0.31 & 6075 & 32.65 & 39.36 & 2.15&1.21 \\
0.36 & 0.51 & 0.15 & 0.43 & 5079 & 42.25 & 45.42 & 2.51&1.08 \\
0.51 & 0.73 & 0.22 & 0.61 & 3479 & 47.74 & 46.36 & 2.66&0.97 \\
0.73 & 1.03 & 0.30 & 0.87 & 1983 & 48.05 & 45.28 & 2.83&0.95\\
1.03 & 1.45 & 0.42 & 1.22 & 1119 & 45.67 & 46.73 & 3.35&1.04 \\
1.45 & 2.06 & 0.61 & 1.73 &  658 & 43.98 & 43.71 & 3.61&1.00 \\
2.06 & 2.91 & 0.85 & 2.45 &  401 & 45.94 & 44.80 & 4.44&0.99 \\
2.91 & 4.11 & 1.20 & 3.46 &  273 & 52.53 & 51.57 & 6.02&1.00 \\
4.11 & 5.82 & 1.71 & 4.89 &  176 & 56.53 & 56.02 & 7.90&1.02 \\
5.82 & 8.23 & 2.41 & 6.92 &  152 & 82.51 & 78.73 &11.57&0.98 \\
8.23 & 11.60& 3.37 & 9.77 &  122 & 112.16& 112.62&18.17&0.98 \\
11.60& 23.30&11.70 &16.44 &  175 & 170.17& 171.87&23.93&1.01 \\ 
\hline
\end{tabular}
\tablefoot{Columns are as follows: $S_{\rm min}$ and $S_{\rm max}$ are the minimum and maximum values of the flux density interval (in mJy), respectively; $\Delta$S is the flux density interval (in mJy); x is the geometric mean of $S_{\rm min}$ and $S_{\rm max}$; N$_S$ is the  number of observed radio sources in each bin; $n$ and $n_{\rm corr}$ denote the normalized source counts (in Jy$^{3/2}$\,sr$^{-1}$) obtained from $N_S$ before and after correcting for the false detections and completeness factors, respectively; err$_{\rm y}$ is an estimate of the errors associated to the corrected differential source counts (in Jy$^{3/2}$\,sr$^{-1}$) that takes into account both the Poissonian contribution and the deviations produced by different simulations; $C_{\rm sim}$ is the average correction factor derived from the simulations.}
\label{tab:counts}
\end{table*}

\subsubsection{Mock radio sources catalogue: injection and recovering}
We generated three different catalogues of mock radio sources using the flux density and angular size distributions as described above.
The mock sources were inserted, as circular Gaussian components, in the residual image and PyBDSF was run with the same parameters used for the real image to recover the mock sources.
The catalogues of retrieved mock sources with SNR$\ge 5$ were cross matched with their respective input catalogues producing the recovered mock sources matched catalogues (simply referred as matched catalogues in the following). We then used the method discussed in Sec.~\ref{sec:resunres}
to classify the mock sources in the matched catalogues as resolved or unresolved. The mock sources are not affected by residual calibration errors or bandwidth smearing and, therefore, the $S_T/S_p$ versus SNR diagram shows no systematic offset. This means A=1 in eqs. (2) and (3). For the B parameter in eqs. (2) and (3) we use B=3.0, the same value adopted for the real sources. 
Summarising, for the matched catalogues we use the same selection function used for the real sources (after the correction for the radially dependent offset). The percentage of resolved mock sources in the matched catalogue is 23\%, very well consistent with the 24\% we have found for the real sources.

Finally, the mock sources are split into different bins of flux density. For the input catalogues, the flux density is that generated using the input flux density distribution, whilst for the matched catalogue
the flux density is the measured flux density for the mock sources classified as resolved or the peak flux density for the unresolved ones. The ratio, for each flux density bin, between the number of sources in input and the number of retrieved sources in the same flux density bin is the correction factor that needs to be applied to the source number counts.
We find good consistency, within $1\sigma$, in the values of the correction factors among the three different catalogues of mock sources and therefore we adopted an average correction factor for each flux density bin. 
The final values of the correction factors are reported in Tab.~\ref{tab:counts}.

As discussed in Sec.~\ref{sec:resunres} the method adopted
to classify a source as resolved or unresolved is an approximation and somewhat arbitrary. Therefore, we used also
slightly different B values and/or shape of the selection function (Eq. 2) to derive alternative versions of the matched catalogues for the mock radio sources. Then, we compared the corrections factors we obtained for these catalogues with those originally derived. The differences in the correction factors obtained from different selection functions are taken into account in an additional fractional error quantified as $\sim 5\%$.

\section{Differential source counts}
\label{sec:counts}
The Euclidean normalized source counts at 144 MHz obtained for the EDFN are presented in Fig.~\ref{fig:counts} and listed in Tab.~\ref{tab:counts}.
The errors associated to the number counts contain the term due to the Poisson statistics for the number of sources in each flux density bin, plus the fixed term of 5\% as discussed above.
For the purposes of this paper it is useful to compare the source counts derived from the EDFN with those derived from the other LoTSS Deep Fields \citep{2021A&A...648A...5M}, also shown in Fig.~\ref{fig:counts},  since we used a different method to assemble the final source catalogue and to derive the counts' corrections. A more comprehensive analysis on the source counts is postponed  until the full set of observations of the EDFN is completed and processed.
The lower panel of Fig.~\ref{fig:counts} is a zoom of the region between 0.2 mJy and 2 mJy at 144 MHz (corresponding to about 0.04 mJy and 0.4 mJy at 1.4 GHz assuming a spectral index $\alpha=0.7$),  where it is quite common to
find significant differences in the source counts from different surveys at the depths we are achieving. Such differences are usually interpreted as a mixing of cosmic variance and uncertainty in the derived correction factors. In this panel, for comparison, we also show the uncorrected normalized source counts (empty circles with no errors).

The source counts obtained for the EDFN are in good agreement with those derived from the other LoTSS Deep Fields
and this is reassuring considering the differences in obtaining them. 
One major difference between the source counts derived from the EDFN and the other LoTSS Deep Fields is the method applied to derive the correction factors for completeness. For the EDFN we produced mock catalogues of radio sources that were added to the residual radio image and then retrieved with the same procedure adopted for the real ones, while for the other LoTSS Deep Fields
\citet{2021A&A...648A...5M} derived the correction factors using a more theoretical approach.

In addition, the counts for Elais N-1, Lockman Hole and Bootes have been obtained using a sample of optically identified and fully deblended radio sources, while for the EDFN only a ``first order'' deblending was performed using the unWISE catalogue. 
Most of the sources that went through deblending had
fluxes around $\sim 1$-2 mJy and the higher value of the source counts in this flux density range suggests that it is possible that not all the sources, that needed to be deblended, were actually deblended. It is worth noting that the number of these potentially not deblended sources is rather small: about 50 sources that were not deblended can explain the somewhat higher value of the counts in the EDFN around $\sim 1$-2 mJy. Clearly, the same effect would lead to an underestimate of the source counts at fluxes $\lsim 1$ mJy producing the sort of discontinuity in the
shape of the EDFN source counts around 1 mJy.
This effect was indeed noted in \citet{2021A&A...648A...5M} when comparing source counts obtained
from raw catalogues (containing radio sources which were not optically identified and not checked and corrected for deblending) and final catalogues (containing only optically identified radio sources with source deblending when needed).
Summarising, the EDFN source counts are generally between those derived from the ELAIS-N1 field, that are systematically slightly lower than the other two deep fields, and those from the Lockman Hole and Bootes fields. The observed field-to-field differences in the source counts at sub-mJy levels are typically around a few percent
and these can be justified both by the different methods adopted to derive the source counts (e.g. EDFN with respect to other LoTSS deep fields) and the expectations from sample variance for surveys 
covering areas $\lsim 10$ deg$^2$ \citep{2013MNRAS.432.2625H,2018MNRAS.481.4548P}.

Once the LoTSS Deep Field project will be completed and fully analysed, it will deliver images and catalogues of radio sources
down to $\sim 60\,\mu$Jy\,beam$^{-1}$ ($5\sigma$) at 144 MHz (corresponding to $\sim 10\,\mu$Jy\,beam$^{-1}$ at 1.4 GHz)
over a 
region of $\sim 50$ deg$^2$. Considering the range of angular resolutions that LOFAR with the international stations can probe, as well as the multi-wavelengths ancillary observations already available or that will be soon available for these fields, in the next few years it is reasonable the LoTSS Deep Fields will allow us to make a major leap in our comprehension of the processes related to galaxy and black-hole formation and co-evolution over cosmic time from a radio perspective.

\section{Summary}
In this paper we present the image and catalogue derived from the first 72 hours of LOFAR observations at 144 MHz covering the central 10\,deg$^2$ region of the Euclid Deep Field North. The image has an angular resolution of $6\asec$ and a central r.m.s. sensitivity of $32\, \mu$Jy\,beam$^{-1}$. 

The main results presented in this paper can be summarised as follows:
\begin{itemize}
\item
We compared images obtained from the same data sets but produced 
with different faceting geometry and starting sky models in the direction-dependent calibration pipeline (\texttt{\small DDF-pipeline}).
We found that for sources within $r<2$ deg from the field centre, the peak or  total flux density are usually consistent within $\lsim 5\%$, while sources at $r>2$ deg can have larger differences up to $\sim$ 10\%-15\%  for sources at $r>3$ deg.
Whilst the two images were obtained running \texttt{\small DDF-pipeline}
on different computers with different individual software package versions, the radial dependency  of the dispersion of the measured flux ratios in the two images suggests that a different faceting geometry could be responsible for the observed trend.

\item
From the inner 10 deg$^2$ circular region ($r<1.784$ deg) we derived a $5\sigma$ catalogue listing about 23,000 radio sources.
We used the unWISE 5 year catalogue to check for possible blended sources in all the objects that were classified as multiple or
complex components by PyBDSF. We also visually inspected the extended
sources to properly assign radio lobes and diffuse components to the correct radio source.

\item
We performed a detailed analysis of the properties of the derived catalogue of radio sources finding a radial dependent effect on the $S_T/S_p$ ratio. We interpreted this as an underestimate of the peak brightness with increasing distance from the field centre as due to residual direction-dependent calibration errors. For the purpose of deriving the 
radio source counts we corrected the peak brightness of the sources classified as unresolved.

\item
Samples of mock sources following realistic flux density and angular size distributions were used to derive the completeness correction factors to be applied to the observed source counts. An estimate of the source false detection rate was obtained as well.  We found  $\sim 3\%$
of false detection rate for sources in the range $5.0\le {\rm SNR} <5.5$ and about $\sim 1.5\%$ in the range $5.5\le {\rm SNR} <6.0$. For sources with
SNR $>10$ the false detection rate is $<0.5\%$ and practically negligible.

\item
The final source counts obtained in the EDFN are consistent with those obtained for the other LoTSS Deep Fields (ELAIS-N1, Lockman Hole and Bootes) and, in particular, they lay between those derived from ELAIS-N1 and those obtained for the other two fields.
In our analysis we also explored the contribution to the errors associated to the source counts produced by slightly different selection functions to separate resolved from unresolved sources. Such differences can be 
quantified in a 5\% additional noise factor that we factored into the final errors shown in Fig.\,\ref{fig:counts} and listed in Tab.\,\ref{tab:counts}.
\end{itemize}

\begin{acknowledgements}
MBond, IP, MBona, MBr, MM, LP, RS  acknowledge support from INAF under the Large Grant 2022 funding scheme (project "MeerKAT and LOFAR Team up: a Unique Radio Window on Galaxy/AGN co-Evolution”. PNB is grateful for support from the UK STFC via grant ST/V000594/1.
MBri acknowledges support from the ERC-Stg "DRANOEL", no. 714245, from the agreement ASI-INAF n. 2017-14-H.O and from the PRIN MIUR 2017PH3WAT "Blackout".
MA acknowledges support from the VENI research programme with project number 202.143, which is financed by the Netherlands Organisation for Scientific Research (NWO).
LKM is grateful for support from the Medical Research Council [MR/T042842/1]. 
RJvW acknowledges support from the ERC Starting Grant ClusterWeb 804208. 
GJW gratefully acknowledges an Emeritus Fellowship from The Leverhulme Trust.
This paper is based on data obtained with the International LOFAR Telescope (ILT) under project codes LC12\_027.
AB acknowledges financial support from the European Union - Next Generation EU.

LOFAR is the Low Frequency Array, designed and constructed by ASTRON. It has observing, data processing, and data storage
facilities in several countries, which are owned by various parties (each with their own funding sources), and which are collectively operated by the ILT foundation under a joint scientific policy. The ILT resources have benefited from the following recent major funding sources: CNRS-INSU, Observatoire de Paris and Université d’Orléans, France; BMBF, MIWF-NRW, MPG, Germany; Science Foundation Ireland (SFI), Department of Business, Enterprise and Innovation (DBEI), Ireland; NWO, The Netherlands; The Science and Technology Facilities Council, UK; Ministry of Science and Higher Education, Poland; The Istituto Nazionale di Astrofisica (INAF), Italy.

This research made use of the OCCAM supercomputing facility 
run by the Competence Centre for Scientific Computing an interdepartmental advanced research centre, that specializes in High Performance Computing (HPC).

The unWISE coadded images and catalogue are based on data products from the Wide-field Infrared Survey Explorer, which is a joint project of the University of California, Los Angeles, and the Jet Propulsion Laboratory/California Institute of Technology, and NEOWISE, which is a project of the Jet Propulsion Laboratory/California Institute of Technology. WISE and NEOWISE are funded by the National Aeronautics and Space Administration. 

For the purpose of open access, the author has applied a Creative Commons Attribution (CC BY) licence to any Author Accepted Manuscript version arising from this submission.
 
\end{acknowledgements}

\bibliographystyle{aa} 
\bibliography{paper-main.bib} 

\end{document}